\documentstyle[aps,epsf]{revtex}

\newcommand{\tauc}{\tau_{\scriptscriptstyle T}}
\newcommand{\tauh}{\tau_{\scriptscriptstyle H}}
\newcommand{\kat}{\kappa_{\scriptscriptstyle T}}
\newcommand{\katp}{\kappa'_{\scriptscriptstyle T}}
\newcommand{\cv}{c_{\scriptstyle x}}

\newcommand{\leqa}{\stackrel{<}{\scriptstyle \sim}}
\newcommand{\geqa}{\stackrel{>}{\scriptstyle \sim}}
\newcommand{\kf}{k_{\scriptscriptstyle F}}
\newcommand{\kb}{k_{\scriptscriptstyle B}}
\newcommand{\ef}{E_{\scriptscriptstyle F}}
\newcommand{\xh}{x_{\scriptscriptstyle H}}

\begin{document}

\draft 
\title {Thermodynamics of small Fermi systems: quantum statistical
fluctuations} 
\author{\large P. Leboeuf and A. G. Monastra}
\address{Laboratoire de Physique Th\'eorique et Mod\`eles Statistiques
\footnote{Unit\'e de recherche de l'Universit\'e de Paris XI associ\'ee au 
  CNRS}, B\^at. 100, \\ 91405 Orsay Cedex, France}
\maketitle

\begin{abstract}
We investigate the probability distribution of the quantum fluctuations of
thermodynamic functions of finite, ballistic, phase--coherent Fermi gases.
Depending on the chaotic or integrable nature of the underlying classical
dynamics, on the thermodynamic function considered, and on temperature, we
find that the probability distributions are dominated either (i) by the local
fluctuations of the single--particle spectrum on the scale of the mean level
spacing, or (ii) by the long--range modulations of that spectrum produced by
the short periodic orbits. In case (i) the probability distributions
are computed using the appropriate local universality class, uncorrelated
levels for integrable systems and random matrix theory for chaotic ones. In
case (ii) all the moments of the distributions can be explicitly computed in
terms of periodic orbit theory, and are system--dependent, non--universal,
functions. The dependence on temperature and number of particles of the
fluctuations is explicitly computed in all cases, and the different relevant
energy scales are displayed.
\end{abstract}

\date{}

\section{Introduction} \label{1}

Self-consistent theories which admit almost independent motion of
quasiparticles and give rise to Fermi gases constitute a basic tool to
describe fermionic many-body systems. Well known examples are
independent-particle models in nuclear physics, quantum dots and wires in
condensed matter physics and, within a mean-field theory for the valence
electrons, the jellium model for the electronic properties of metals and of
simple alkali metal clusters. One of the most spectacular predictions of
mean-field theories is the occurrence of shell effects in the energy of the
fermion gas as a function of the number of particles, that leads to magic
numbers and deformations in nuclei and metallic particles \cite{bm,brack}. The
shell and super--shell structures are quantum fluctuations that describe the
bunching of the single-particle energy levels with respect to their average
behavior. From a semiclassical point of view \cite{bb,sm}, the bunching of the
energy levels is interpreted as a modulation produced by the classical
periodic orbits.

Similar quantum fluctuations are present in all the equilibrium thermodynamic
functions describing the many-body system. As we will see in detail in the
following, there are two main features of the single--particle spectrum that
influence the thermodynamic quantum fluctuations of a Fermi gas. On the one
hand, the spectral fluctuations on local scales, i.e. in the range of the mean
level spacing $\delta$ between single--particle energy levels. These local
fluctuations are universal, and the universality class depends on the nature
of the single-particle dynamics \cite{bgs,bt2}. There are two main classes,
corresponding to regular and irregular (fully chaotic) motion, respectively
(with subclasses in the latter case depending on the presence or not of time
reversal symmetry). The local statistics of regular systems are described by a
set of uncorrelated levels, while the chaotic ones by random matrix ensembles.

The second ingredient that strongly influences the thermodynamic fluctuations
are the shell effects, i.e. the long range modulations of the spectrum
produced by the short periodic orbits. These occur on a scale $E_c$ related to
the inverse time of flight across the system, with typically $E_c \gg \delta$.
Contrary to the local fluctuations, the long--range modulations are
system-dependent, since the short orbits are. The amplitude of this effect
depends on the nature of the dynamics. If the classical motion is
regular, symmetries are present, periodic orbits come in degenerate families,
and shell effects are important. On the contrary, if the corresponding
classical motion is chaotic, the periodic orbits are isolated and the
long-range modulations will have, compared to the regular case, a smaller
amplitude. Different aspects of both quantum phenomena, namely the
universality of the single--particle fluctuations on a scale $\delta$ and the
appearance of long--range shell effects on scales $E_c$, and their influence
in the thermodynamics of Fermi gases, have been considered in nuclear and
atomic physics, and in different condensed--matter systems (see for example
Refs.~\cite{bm,brack,efetov,imry}). In most cases the interest concentrated on
particular systems, or in the study of mean properties.

Our main goal is, within a Fermi gas approximation, to provide a statistical
analysis of the quantum fluctuations of the different thermodynamic functions,
like for example of the total energy of the gas, the entropy, etc, and to
explicitly compute their probability distribution function. Special attention
is devoted to the temperature dependence and to the typical magnitude of the
fluctuations. Other aspects, like the autocorrelation functions and the
connexion between the fluctuations of the gas and those of the
single--particle levels at Fermi energy, are also investigated. We concentrate
in the fluctuations of equilibrium properties of {\sl ballistic} Fermi systems
(i.e. with no disorder present in the potential) whose mean-field single
particle motion is either regular or fully chaotic. Schematically we have in
mind a mean-field potential which is almost flat in the interior and rises at
the surface, like the Woods-Saxon potential in nuclear physics, or the static
homogeneous potential produced by the ionic background in the jellium model.
Under these conditions, and in particular in the extreme case of a sharp
boundary, the nature of the classical dynamics is determined by the shape of
the cavity and by the external fields acting on the system. We will ignore
physical effects related to the spin degree of freedom, and we will moreover
assume that the inelastic processes can be ignored, namely the coherence
length of the fermions is much larger than the system size.

The picture that emerges from our analysis is very rich. The quantum
thermodynamic fluctuations can be either dominated by the local universal or
by the long--range system--specific fluctuations of the single particle
spectrum. Among these two possibilities, the prevailing aspect depends on the
quantity considered, on the nature of the underlying classical dynamics, and
on temperature. When the fluctuations are dominated by the local fluctuations,
the corresponding thermodynamic probability distributions are universal. When
conveniently normalized, they only depend on the temperature and the quantity
considered, but not on any specific property of the system (aside the regular
or chaotic nature of the single--particle motion). The universality of the
thermodynamic fluctuations follows the classification of the local
fluctuations given above: one universality class for systems whose classical
single--particle dynamics is regular, another for those with a chaotic
dynamics (with subclasses depending on the presence or not of time reversal
symmetry). Universal fluctuations are observed, for some thermodynamic
functions, at temperatures much smaller than $E_c$. Examples are the
fluctuations of the number of particles for systems with fixed chemical
potential when the underlying classical dynamics is chaotic, and those of the
entropy and the specific heat for regular and chaotic motion. The
corresponding probability distributions are found to be, in general,
non-Gaussian. For temperatures of order $\delta$ their shape is very sensitive
to temperature variations, but they always remain in the corresponding
universality class. When temperatures of order $E_c$ are reached, the
universality is lost. At those temperatures the short orbits become dominant,
producing system--dependent effects. The moments of the distributions can then
be computed from periodic orbit theory, as described below. At still higher
temperatures the fluctuations are exponentially suppressed.

In contrast to the first type described above, sometimes the thermodynamic
fluctuations are dominated by the long range modulations of the
single--particle spectrum produced by the short periodic orbits. Due to the
system specific nature of the short orbits, the corresponding probability
distributions are non-universal, non-Gaussian functions. We have obtained
explicit convergent expressions to compute all the moments for this type of
fluctuations in terms of periodic orbits. Interference effects between
repetitions of the orbits produce, in particular, non--vanishing odd moments
and lead to asymmetric distributions. The fluctuations of all the
thermodynamic functions are of this type for temperatures of order $E_c$ or
higher. However, some thermodynamic functions have non-universal fluctuations
{\it at any temperature}. Examples of the latter behavior are the fluctuations
of the total energy of the gas, the response of the system when an external
parameter is varied and, only in the case of a regular single--particle
dynamics, the fluctuations of the number of particles for systems with fixed
chemical potential. Due to the dominance of the long range modulations, the
probability distributions are relatively insensitive to temperature variations
up to temperatures of order $E_c$. At temperatures higher than $E_c$ the
typical size of the fluctuations decay exponentially. Finally, the
autocorrelation functions in this case present long range order.

As a general rule, but with exceptions depending on the quantity and on the
temperature considered, clear signatures of the regular or chaotic nature of
the dynamics are found in the fluctuations. For example, their typical size is
in general much larger in integrable systems compared to chaotic ones. The
relative amplifying factor is found to be some growing function of the
adimensional parameter $g = E_c /\delta$. Besides, and irrespective of the
nature of the dynamics, the fluctuations of the gas are always much larger
than the corresponding fluctuations of the single--particle energy levels at
Fermi energy. Their relation is also expressed in terms of $g$.

The paper is organized as follows. Section \ref{2} contains the basic
definitions concerning the thermodynamics of a Fermi gas. An arbitrary
thermodynamic function is decomposed into a smooth plus an oscillatory part.
The smooth part describes the usual thermodynamics of the gas (bulk
properties), while the oscillatory part are the quantum finite-size
fluctuations. The latter are expressed in terms of the periodic orbits of the
classical motion associated with the corresponding single-particle dynamics.
The validity of this approximation and the main energy scales involved in the
problem are also introduced. In Section \ref{3} the variance of the
thermodynamic functions is written in terms of the spectral form factor, i.e.
the Fourier transform of the single--particle spectral two-point correlation
function (Eq.(\ref{grandk})). This connexion allows to make a qualitative
analysis of the behavior of the moments of the probability distribution of the
different functions and for the different kinds of classical motions. The
second moment is explicitly computed in Section \ref{4} using a schematic
approximation for the short-time dynamics that provides simple but
qualitatively useful expressions. The dependence on particle number and on
temperature is explicitly considered. The complete characterization of the
probability distribution of the fluctuations is included in Section \ref{5}.
The moments are explicitly computed for quantities whose distribution is
dominated, by the short orbits. On the contrary, local uncorrelated levels and
random matrix fluctuations are used to compute the distribution of universal
functions (at low temperatures). Similar arguments are exploited in Section
\ref{6} to compute the autocorrelation of the thermodynamic functions as a
function of the different parameters. The application of our general results
to different systems, like the atomic nuclei, mesoscopic quantum dots and
metallic clusters is briefly discussed in Section \ref{7}. This section
contains also a discussion on the relation between the fluctuations of the gas
and those of the individual single--particle energy levels. Our final remarks
are included in Section \ref{8}. All over the paper we compare some of the
results with numerical simulations of Fermi gases contained in
regular and chaotic cavities.

\section{General setting} \label{2}

\subsection{Basic equations} \label{2a}

When the chemical potential $\mu$ is fixed the thermodynamic behavior of a
non--interacting Fermi gas is described by the grand potential
\begin{equation} \label{grand}
\Omega (x,\mu,T)= - \kb T \int dE ~\rho(E,x) ~ \log [1+e^{(\mu-E)/\kb T}] \ , 
\end{equation}
where
\begin{equation} \label{rho}
\rho (E,x) = g_s \sum_j \delta \left[ E - E_j(x) \right]
\end{equation}
is the density of states of the single--particle energy levels $E_j(x)$. The
parameter $x$ denotes the dependence on some external parameter, $T$ is the
temperature, and $\kb$ is Boltzmann's constant. The prefactor $g_s=2$ when
spin degeneracy exists. For simplicity we will from now on omit it, but it can
be easily restored in the final results.

When the number of particles $N$ in the gas is fixed, the
energy of the system is
\begin{equation} \label{energy}
U (x,N,T)= \int dE~E~\rho(E,x)~f(E,\mu,T) \ , 
\end{equation}
where $\mu$ is determined from the equation
\begin{equation} \label{nmu}
N = \int dE~\rho(E,x)~f(E,\mu,T) \ , 
\end{equation}
and $f$ is the Fermi function
\begin{equation} \label{f}
f(E,\mu,T) = \frac{1}{1+e^{(E-\mu)/\kb T}} \ .
\end{equation}

Knowing these two thermodynamic potentials, other thermodynamic functions are
directly obtained by differentiation. For example, in the grand canonical
formalism the expectation value of the number of particles in the system is
\begin{equation}
{\cal N} (x,\mu,T)= - \left. \frac{\partial \Omega}{\partial \mu} \right|
_{x,T} \ . \label{n}
\end{equation}
Another important thermodynamic quantity is the reaction of the system to a
variation of the external parameter $x$, characterized by the response
function (here in the grand canonical formalism)
\begin{equation}
R(x,\mu,T)= - \left. \frac{\partial \Omega}{\partial x} \right| _{\mu,T} \ .
\label{r}
\end{equation}
The physical interpretation of $R$ depends on the sample geometry and on the
nature of the parameter $x$. $R$ is the orbital magnetization when $x$ is a
magnetic field, a persistent current when the sample geometry is annular and
$x$ is an Aharonov-Bohm flux, or a force (pressure) when $x$ controls the
shape of the confining potential (like when deforming a nucleus for example;
at equilibrium, the nucleus shape is defined by $R=0$). Finally, we will also
consider the entropy
\begin{equation}\label{s}
S (x,\mu,T)= - \left. \frac{\partial \Omega}{\partial T} \right| _{\mu,x}
\end{equation}
from which, in particular, the many--body level density may be computed by
exponentiation, and the specific heat (in the canonical ensemble)
\begin{equation}\label{cv}
\cv = \left. \frac{\partial U}{\partial T} \right|_{N,x} \ .
\end{equation}
The case of the susceptibility, i.e. the variation of the response function 
(the second derivative of the grand potential)
\begin{equation}\label{curv}
\chi (x,\mu,T)= \left. \frac{\partial R}{\partial x} \right| _{\mu,T} = -
\left. \frac{\partial^2 \Omega}{\partial x^2} \right| _{\mu,T} 
\end{equation}
is a more subtle problem and will not be considered here in detail.

Assuming the single-particle spectrum Eq.~(\ref{rho}) known for any $x$, the
integration in Eq.(\ref{grand}) is straightforward. It leads, by
differentiation, to sums over the single-particle energy levels whose
structure depends on the quantity considered. As the temperature is lowered
the Fermi factor $f(E,\mu,T)$ involved in the sums tends to the Heaviside step
function, which restricts the sums to the single--particle energy levels
satisfying the condition $E_j (x) \leq \mu$. In this limit the grand potential
is closely related to the ground state energy of the Fermi gas (it is the sum
of the single particle energy levels measured with respect to the Fermi
energy), ${\cal N}$ is the number of occupied single--particle levels, and the
response of the system is the sum up to the Fermi energy of the
single-particle contributions $\partial E_j/\partial x$. They all involve the
ensemble of particles of the Fermi gas. In contrast, other quantities, like
the entropy and the specific heat, vanish at zero temperature.

\subsection{Semiclassical approximation} \label{2b}

We now express the thermodynamic functions in terms of classical quantities
related to the single--particle dynamics (cf Ref.\cite{ruj} for a good
introduction). The density of states in Eq.(\ref{grand}) is written as a sum
of smooth plus oscillatory terms
\begin{equation}
{\rho} (E,x) = \overline{{\rho}} (E,x) + \widetilde{{\rho}} (E,x) \ .
\end{equation}
This approximation is valid in the semiclassical regime $\mu \gg \delta$ where
the typical wavelength of the fermions at energy $\mu$ is much smaller than
the system size, and fails at the bottom of the spectrum. The first, smooth
term $\overline{\rho}$ corresponds to the bulk contribution. For example, when
the gas is simply contained in a finite volume $V$ of space having an
arbitrary shape, and the single--particle levels are filled up to a Fermi
energy $\ef$ with particles of mass $m$, the average number of occupied levels
is, to leading order
\begin{equation}\label{nb}
{\overline {\cal N}} = \frac{4 V}{3 \sqrt{\pi}} \left( \frac{m}{2 \pi \hbar^2}
\right)^{3/2} \ef^{3/2} \ ,
\end{equation}
while the mean level density ${\bar \rho}$ and corresponding mean level
spacing $\delta$ between single--particle states are
\begin{equation}\label{mls}
{\bar \rho} = \delta^{-1} = \frac{2 V}{\sqrt{\pi}} \left( \frac{m}{2 \pi \hbar^2} \right)^{3/2} \ef^{1/2} \ .
\end{equation}
The substitution in Eqs.(\ref{grand}) and (\ref{energy}) of the density of
states by its average behavior, Eq.(\ref{mls}), gives rise to the smooth, bulk
classical expressions for the thermodynamics of the gas \cite{landau}. In
finite fermionic systems, like the atomic nucleus or a metallic particle, the
smooth part coming from the single--particle potential is of no use, and is
rather computed in a self--consistent manner from the original many--body
problem \cite{bdjpsw}.

The most relevant contribution to the many--body problem is the fluctuating
part, which is of quantum mechanical origin, and describes the discreteness of
the single--particle spectrum. It is given by a sum over the classical
periodic orbits \cite{bb,gutz}. To leading order in an $\hbar$--expansion
\begin{equation}
\widetilde{{\rho}} (E,x)= 2 \sum_p \sum_{r=1}^{\infty} A_{p,r} (E,x) \cos
\left[ r S_p(E,x)/\hbar+\nu_{p,r} \right] \ . \label{rhosc}
\end{equation}
The sum is over all the primitive periodic orbits $p$ (and its repetitions
$r$). The orbits are characterized by their action $S_p$, period $\tau_p=d
S_p/dE$, stability amplitude $A_{p,r}$, and Maslov index $\nu_{p,r}$.

Explicit expressions for $A_{p,r}$ are available. Their functional form
depends on the nature of the dynamics. For a $d$--dimensional integrable
system, where action-angle variables exist, the periodic orbits are organized
on resonant tori, i.e. $d$--dimensional manifolds having the topology of a
torus with commensurate frequencies. If $I_j$ are the action variables, then
the frequencies are $\omega_j=\partial H/\partial I_j= 2\pi m_j/\tau_p$, where
the $m_j$ are integers that label the periodic orbits (that we abbreviate in
the unique index $p$). When the $\{ m_j \}$ have no common divisor they label
a primitive periodic orbit, otherwise it is a repetition. The $A_p$'s take in
this case the form \cite{bt1}
\begin{equation}
A_p^2 = \frac{(2 \pi)^{d-1}}{\hbar^{d+1} ~ \tau_p^d ~ \left|\det \{
\partial \omega_j/\partial I_k \}_p ~ \sum_j \omega_j \cdot \partial
I_j/\partial \tau_p \right|} \ . \label{berry-tab}
\end{equation}
The repetitions do not appear explicitly, and are considered as primitive
periodic orbits. Instead, in chaotic system the amplitudes are \cite{gutz}
\begin{equation}
A_{p,r} = \frac{\tau_p}{h \sqrt{\left|\det \left( M_p^r-I \right)
\right|}} \ .
\end{equation}
Here $I$ is the identity matrix and $M_p$ the monodromy matrix obtained from
the linearization of the equations of motion in the vicinity of the
corresponding primitive periodic orbit. In both cases the Maslov indices
$\nu_{p,r}$ are constant phase factors (locally) independent of the energy
and of the external parameter.

Eq.(\ref{rhosc}) is inserted in Eq.(\ref{grand}) to compute the oscillatory
part of the grand potential. To leading order in $\hbar$ and for low
temperatures ($\kb T \ll \mu$, degenerate gas approximation)
\begin{equation}
\widetilde{\Omega} (x,\mu,T) \approx 2 \hbar^2 \sum_p \sum_{r=1}^{\infty}
\frac{A_{p,r} ~ \kat(r ~ \tau_p)}{r^2 ~ \tau_p^2} \cos(r
S_p/\hbar+\nu_{p,r}) \label{grandosc} \ .
\end{equation}
The classical functions entering in this expressions are evaluated at $E=\mu$,
and they all depend on the external parameter $x$. The temperature introduces
the prefactor $\kat(\tau)$, which acts as an exponential cut off for the long
orbits \cite{bm}
\begin{equation}\label{kat}
\kat(\tau) = \frac{\tau / \tau_{\scriptscriptstyle T}}
{\sinh (\tau / \tauc)} \ , \;\;\;\;\;\;\;\;\;\;\;\;\; \tauc = h/ (2\pi^2 \kb T)
\ .
\end{equation}
For temperatures such that $\tauc \ll \tau_{min}$, with $\tau_{min}$ the period
of the shortest periodic orbit, the quantum fluctuations are washed out and
only the smooth behavior given by ${\bar \rho}$ survives.

Similarly we may analyze the energy of the gas at a fixed number of particles.
The smooth part of $U$ differs from that of $\Omega$, but to leading order in
a semiclassical expansion we find that their fluctuating parts coincide.
$\widetilde{U}$ is thus given by an equation identical to
Eq.~(\ref{grandosc}), with $\mu$ not a continuous variable but now a function
of $N$, $x$ and $T$ determined by Eq.(\ref{nmu}) but where ${\bar \rho}$ is
used instead of $\rho$,
\begin{equation}\label{ut}
\widetilde{U} (x, N, T) = \widetilde{\Omega} (x, \mu(x, N, T), T) \ .
\end{equation}

The prefactor $A_{p,r} ~ \kat(r ~ \tau_p)/ \tau_p^2$ in Eq.~(\ref{grandosc})
varies smoothly with the external parameter, the chemical potential, or
temperature. The main contribution to the oscillatory behavior of the grand
potential comes in fact from the variations of the phase factor $r S_p (x,\mu)
/\hbar$ with respect to $x$ and $\mu$. When differentiating
Eq.~(\ref{grandosc}) or Eq.~(\ref{ut}) to compute the fluctuating part of
other thermodynamic quantities according to Eqs.~(\ref{n})-(\ref{curv}) we
only keep, to first order in a semiclassical expansion in $\hbar$, the terms
coming from the variations of the phase factor. In this approximation, the
fluctuating part of all the thermodynamic quantities considered has the same
structure, and can be written in a compact form as
\begin{equation}
\widetilde{\Phi} (x,\mu,T) = 2 ~ {\cal C} \sum_p \sum_{r=1}^{\infty} {\cal
A}_{p,r} (x,\mu,T) \cos \left[ r S_p(x,\mu)/\hbar+\nu_{p,r} \right] \ .
\label{genosc}
\end{equation}
The constant coefficient ${\cal C}$ includes all the terms not depending on
the periodic orbits. This coefficient, as well as the orbit-dependent prefactor
${\cal A}_{p,r}$ computed in the leading semiclassical approximation, are
given for some of the thermodynamic functions in Table~I. When necessary, the
sine functions have been systematically transform into cosines by subtracting
a $\pi/2$ to each Maslov index. The prime in $\katp$ denotes the derivative of
$\kat$ with respect to temperature, $\katp = \partial \kat/\partial T$, while
\begin{equation} \label{qp}
Q_p = \left. \frac{\partial S_p}{\partial x} \right|_\mu \ .
\end{equation}

Eq.(\ref{genosc}), together with Table~I, are the basic expressions on which
the study of the probability distributions of the fluctuations is based on.
The knowledge of the periodic orbits (or at least of those satisfying, at
finite temperature, $\tau_p \leqa \tauc)$ allows to explicitly compute, using
Eq.(\ref{genosc}), the quantum oscillatory contributions to $\Phi$. We are
interested, however, in a statistical analysis valid for a generic system
rather than in an explicit computation for a particular one.

\vspace{0.5cm}

\begin{center}
\begin{tabular}{||c||c|c|c||}
\hline\hline & & & \\ ~~~ & ${\cal C}$ & ${\cal A}_{p,r}$ & $F(\tau)$ \\
\hline & & & \\ 
$\widetilde{U}$ & $\displaystyle \hbar^2$ & $\displaystyle
A_{p,r} \ \kat (r \tau_p) /r^2 \tau_p^2$ & $\kat^2 (\tau)$ \\ 
\hline & & & \\
$ \widetilde{\cal N}$ & $\hbar$ & $\displaystyle A_{p,r} \ \kat (r \tau_p) /r
\tau_p$ & $\tau^2 \ \kat^2 (\tau)$ \\ 
\hline & & & \\ 
~~~ $ \widetilde{R}$ ~~~ & ~~~ $\hbar$ ~~~ & ~~~ $\displaystyle A_{p,r} \ Q_p
 \ \kat (r \tau_p) /r \tau_p^2$ ~~~ & ~~~ $Q^2 (\tau) \ \kat^2 (\tau)$ ~~~ \\
\hline & & & \\
$\widetilde{\cal S}$ & $\displaystyle \hbar^2$ & $\displaystyle A_{p,r} \
\katp (r \tau_p) /r^2 \tau_p^2$ & $\katp \ \!\!\!^2 (\tau)$ \\ 
\hline\hline
\end{tabular}
\end{center}
%\vskip 1 truecm 
\noindent {\bf Table~I}: the coefficient ${\cal C}$ and prefactor ${\cal
A}_{p,r}$ of Eq.(\ref{genosc}) for different thermodynamic functions. See the
text for details. In the third column is listed the function $F(\tau)$
appearing in Eq.~(\ref{grandk}).

\vspace{0.5cm}

As $T \rightarrow 0$ the sum (\ref{genosc}) over periodic orbits is in general
divergent. For instance, in fully chaotic systems it diverges for temperatures
$T \leqa \hbar \xi / 2 \pi \kb$ (where $\xi$ is the Lyapounov exponent), and
converges otherwise. Using resummation techniques finite sums may be obtained,
as was done for example for the magnetic susceptibility of two-dimensional
billiards in \cite{agam}. However, we will not need to proceed along these
lines since, as we will see, {\sl the moments} of the thermodynamic quantities
based on these sums are, in some cases, {\sl convergent}. In the cases where
they don't, the distributions will be directly determined from the
local universal fluctuations.

\subsection{Averages, energy and parameter scales} \label{2c}

There are several different energy scales that are relevant when considering
the quantum thermodynamic fluctuations. The smallest energy scale in the
problem is the single-particle mean level spacing $\delta = {\bar \rho}^{-1}$
at the Fermi energy $\mu$ (since $\kb T \ll \mu$, we don't make in this
qualitative considerations the difference between the chemical potential and
the Fermi energy). The time scale associated to this
energy is the Heisenberg time
\begin{equation}\label{th}
\tauh = h/\delta \ .
\end{equation}
The largest energy scale is the Fermi energy $\ef$, or chemical potential
$\mu$. The third relevant energy scale has a simple semiclassical origin.
According to Eq.~(\ref{genosc}), locally each periodic orbit $p$ produces a
periodic modulation (in energy) of wavelength $\lambda_p \sim h /\tau_p
(\mu)$. Since there are orbits of arbitrary long period, $\lambda_p$ can be
arbitrarily small. There is on the contrary an upper bound to $\lambda_p$
determined by the period $\tau_{min}$ of the shortest periodic orbit (at
$E=\mu$). The corresponding energy scale is
\begin{equation}\label{ec}
E_c = h/\tau_{min} \ .
\end{equation}
This is the largest scale in which long--range modulations of the
single--particle spectrum occur, and fixes the size of the shell structures at
$E=\ef$. The ratio of $E_c$ to $\delta$ is another important parameter
\begin{equation}\label{g}
g=\tauh/\tau_{min}=E_c/\delta .
\end{equation}
The energy $E_c$ and the adimensional parameter $g$ are, in some respects, the
analog for ballistic systems of the Thouless energy and the adimensional
conductance defined in diffusive systems, respectively \cite{thouless,imry}.
Its physical meaning in ballistic systems is clear from Eq.~(\ref{g}). It is
the number of single--particle states contained in the last shell, which is of
size $E_c$. It is therefore a measure of the collective effect of the
modulations produced by the shell contributions, a ``shell--strength''
parameter. 

Assuming the typical system-size is $L$, it is useful to display the different
energy scales $\delta$, $E_c$, and $\mu$ in terms of the typical number of the
De Broglie wavelengths the system can accommodate at Fermi energy, $\kf L$,
where $\kf$ is the Fermi wavevector. These estimates are based on the
generalization of Eqs.~(\ref{nb}) and (\ref{mls}) to a $d$--dimensional
cavity. On the one hand we have $g = E_c/\delta \propto (\kf L)^{d-1}$.
Moreover, $\mu /E_c = \kf L / 2 \pi$. Thus, in the semiclassical regime $\kf L
\gg 1$ where our description applies, the parameter $g$ is large, and the
different energy scales are well separated, $\delta \ll E_c \ll \mu$.

The fourth relevant energy scale is the temperature. Eq.~(\ref{genosc}) is
valid in the limit $\kb T \ll \mu$ of a strongly degenerate gas, which is the
relevant limit for most physical applications. Notice the large coefficient
$2\pi^2 \approx 20$ involved in the parameter $\tauc$ in Eq.(\ref{kat}), as
compared to Eqs.~(\ref{th}) and (\ref{ec}). This large coefficient cannot be
ignored, and therefore the relevant thermal energy to be compared with the
quantum scales $E_c$ and $\delta$ is $2 \pi^2 \kb T$.

The fluctuating part of a given thermodynamic function $\widetilde{\Phi}$
shows, as a function of the chemical potential $\mu$, oscillations described
by Eq.~(\ref{genosc}). The statistical properties of $\widetilde{\Phi}$, and
in particular its probability distribution, will be computed in a given
interval of size $\Delta \mu$ around $\mu$. This interval must satisfy two
conditions. It must be sufficiently small in order that all the classical
properties of the system remain almost constant. This is fulfilled if $\Delta
\mu \ll \mu$. Moreover, it must contain a sufficiently large number of
oscillations to guarantee the convergence of the statistics. As stated
previously the largest scale associated to the oscillations is $E_c$. Then
clearly we must have $\Delta \mu \gg E_c$. In the semiclassical regime the
hierarchical ordering between the different scales is therefore
$$
\delta \ll E_c \ll \Delta \mu \ll \mu \ .
$$
Since $\mu / E_c \propto \kf L$, a typical scale for the smoothing energy
window is $\Delta \mu/E_c \propto (\kf L)^{1/2}$. We thus define the energy
average of a certain oscillating function as
\begin{equation}
\langle f(\mu) \rangle_{\mu} \equiv \frac{1}{\Delta \mu} \int_{\mu- \Delta
\mu/2}^{\mu+ \Delta \mu/2} f(\mu') d \mu' \ .
\end{equation}

In a similar way a parameter average is defined as
\begin{equation}
\langle f(x) \rangle_{x} \equiv \frac{1}{\Delta x} \int_{x- \Delta x/2}^{x+
\Delta x/2} f(x') d x' \ ,
\end{equation}
where $\Delta x$ is defined by similar arguments as $\Delta \mu$. As in the
energy average, the classical properties must not significantly change in the
interval $\Delta x$, but it must contain several oscillations to make a
statistical analysis appropriate. The largest scale associated to these
parameter oscillations is again related to the shortest orbit, and given by
$h/Q_{min}$.

We finally note that if all these conditions are satisfied both the
energy and parameter averages gives the same result, and consequently commute.

\section{Qualitative analysis of the distributions} \label{3}

By definition, the average value of the fluctuating part $\widetilde{\Phi}$ is
zero. Aside the average value, the variance is the more basic aspect of the
probability distribution of the fluctuations. It provides the typical size of
the oscillations, and can easily be compared with experiments. We will now
compute a general expression for the second moment that allows to make
a qualitative analysis of the distribution of the different thermodynamic
quantities. Similar arguments for the higher moments lead to similar
conclusions.

From Eq.~(\ref{genosc}) the square of $\widetilde{\Phi}$ is expressed as a
double sum over the periodic orbits involving the product of two cosines. The
latter may be expressed as one half the sum of the cosine of the sum and that
of the difference of the actions. The average over the term containing the sum
of the actions vanishes, due to its rapid oscillations on a scale $\Delta
\mu$. Therefore
\begin{equation}
\langle \widetilde{\Phi}^2 \rangle = 2 ~ {\cal C}^2 \left\langle \sum_{p,p'}
{\cal A}_p ~ {\cal A}_{p'} \cos \left( \frac{S_p - S_{p'}}{\hbar} \right)
\right\rangle_{\mu,x} \ . \label{vargen2}
\end{equation}
To simplify the notation we have momentarily considered the repetitions of a
primitive orbit as a different primitive periodic orbit, and have included the
Maslov indices in the definition of the action. Ordering the orbits by their
period, and taking into account the restrictions imposed by the averaging
procedure, we can relate the variance Eq.~(\ref{vargen2}) to the semiclassical
definition of the form factor $K(\tau)$ (i.e., the Fourier transform of the
two-point correlation function $\langle \tilde{\rho} (\mu + \epsilon/2)
\tilde{\rho} (\mu-\epsilon/2) \rangle_\mu$ with respect to $\epsilon$),
expressed as \cite{berry}
\begin{equation} \label{form}
K(\tau)= h^2 \left\langle \sum_{p,p'} A_p ~ A_{p'} \cos \left(
\frac{S_p-S_{p'}}{\hbar} \right) ~ \delta \left[ \tau -
\frac{(\tau_p+\tau_{p'})}{2} \right] \right\rangle _{\mu,x} \ .
\end{equation}
The form factor defined in Eq.~(\ref{form}) has units of time. When
expressed in terms of $K(\tau)$, the variance of the quantum thermodynamic
fluctuations considered above takes the simple form
\begin{equation} \label{grandk}
\langle \widetilde{\Phi}^2 \rangle = \frac{{\cal C}^2}{2 \pi^2 \hbar^2}
\int_0^{\infty} \frac{d\tau}{\tau^4} ~ K(\tau) ~ F (\tau) \ ,
\end{equation}
where $F(\tau)$ depends on the thermodynamic function considered (cf Table~I).
Analogous expressions connecting the variance of different quantities like the
persistent current or the conductance fluctuations have been obtained when
describing disordered metals \cite{montambaux}. In the latter case, the form
factor for times $\tau \ll \tauh$ is closely related to the classical return
probability for a diffusive particle.

To obtain Eq.(\ref{grandk}) we made use of the fact that the orbits giving a
non-zero contribution to (\ref{vargen2}) have similar actions (unless their
average will be zero). This implies that their period is also similar, and can
be considered to be the same in the prefactor (but not in the argument of the
oscillating part). Similar arguments apply to the $Q_p$'s. It can be shown by
arguments invoking the stationarity of the form factor under variations of the
external parameter $x$ that the contributing orbits have very similar
derivatives $Q_p$. The factor $Q^2 (\tau)$ entering the variance of the
response function has a statistical meaning. When replacing the double sum in
Eq.~(\ref{vargen2}) by an integral over the period $\tau$, among all the
periodic orbits $p$ of period between $\tau$ and $\tau + d \tau$ we consider
the distribution of the derivatives $(\ref{qp})$. Then $Q^2 (\tau)$ is the
second moment of that distribution. The expression (\ref{grandk}) for the
response function therefore neglects the non-statistical behavior of the short
orbits.

By definition, the form factor (\ref{form}) is system dependent. But general
statistical statements, depending only on the nature of the underlying
classical dynamics, have been conjectured for ballistic systems. When the
classical dynamics is integrable, the quantum energy levels are believed to
behave as an uncorrelated sequence \cite{bt2}. This implies no
$\tau$-dependence of the form factor,
\begin{equation}\label{kint}
K_{int}(\tau) = \tauh \ ,
\end{equation} 
where $\tauh$ was defined in Eq.~(\ref{th}). The situation is different
for fully chaotic systems, where it has been conjectured that the fluctuations
are described by the corresponding ensembles of random matrix theory
\cite{bgs,bohigas}. For the form factor it gives
\begin{equation} \label{krmt}
K_{rmt} (\tau) = \left\{ \begin{array}{ll} 
\left[ 2 \tau - \tau \log \left( 1 + \frac{2 \tau}{\tauh} \right) \right] 
\Theta (\tauh - \tau) +
\left[ 2 \tauh - \tau \log \left( \displaystyle\frac{2 \tau + \tauh}
{2 \tau - \tauh}\right)\right] 
\Theta (\tau - \tauh)  \;\;\;\;\;\; & \beta = 1 \\ 
\tau \ \Theta (\tauh - \tau) + \tauh \ \Theta (\tau - \tauh)  & \beta = 2 
\ . \\
\end{array}
\right.
\end{equation}
where $\beta=1 ~ (2)$ for systems with (without) time-reversal symmetry (since
we are neglecting the role of spin we do not include the symplectic symmetry).
For short times $\tau \ll \tauh$ this function behaves as
\begin{equation} \label{krmtd}
K_{rmt} (\tau) = \frac{2}{\beta} \ \tau \ .
\end{equation}

Aside the form factor, to compute the variance of the response function
information about the $Q_p$'s is required. For chaotic systems numerical
evidence as well as general considerations suggest a Gaussian distribution,
with zero average (if the area of the cavity is preserved when $x$ varies) and
variance proportional to the period $\tau$ \cite{gsbswz}. Then $Q^2(\tau)
\approx \alpha \tau$, where $\alpha$ is some system-dependent constant which
corresponds to a classical diffusion coefficient \cite{ls}. In the integrable
case no universal distribution exists for the $Q_p$'s, but explicit
calculations for some integrable systems as well as general heuristic
arguments indicate that the second moment is proportional to $\tau^2$
\cite{lm1}. In summary
\begin{equation}\label{qp2} 
Q^2 (\tau) = \left\{ \begin{array}{ll}
\alpha \tau  \;\;\;\;\;\; & \mbox{chaotic} \; , \\ & \\
\omega \tau^2 & \mbox{integrable}  \; . 
\end{array}\right.
\end{equation}

For short times, Eqs.~(\ref{kint}) and (\ref{krmtd}) are good approximations
in real systems in the regime $\tau_{min} \ll \tau \ll \tauh$. Their
semiclassical origin is related to the statistical behavior of the long
classical periodic orbits \cite{ho,berry}.

For times of the order of $\tau_{min}$ no statistical universal behavior of
the form factor exists, and Eq.~(\ref{kint}) and (\ref{krmtd}) do not provide
a good description. For such times the off-diagonal contributions in
Eq.~(\ref{form}) are eliminated by the averaging procedure, and the behavior
of $K(\tau)$ is described by a series of delta peaks at $\tau = \tau_p$
obtained from the diagonal terms $p = p'$ in the double sum
\begin{equation} \label{formd}
K(\tau) \approx h^2  \sum_{p} A_p^2 ~ \delta \left( \tau -
\tau_p \right)  \ .
\end{equation}
The lowest peak is located at $\tau = \tau_{min}$, and for $\tau \leq
\tau_{min}$ the form factor is identically zero.

Based in Eqs.(\ref{grandk})--(\ref{formd}) we now make a simple qualitative
analysis of the variance of the fluctuations of the thermodynamic functions.
To simplify the analysis, in the chaotic case we will use the simple
approximation (\ref{krmtd}) extended to times $\tau \rightarrow \infty$, and
ignore the saturation of the form factor for longer times described by
Eqs.~(\ref{krmt}). By doing this, we are overestimating the weight of the long
orbits. To begin with, we also set the temperature to zero. In Table~I we put
$\kat = 1$ and $\katp = 0$. The variance of the entropy fluctuations therefore
vanishes, as it should. We now discuss the non--vanishing quantities.

Consider for example the variance of the energy (or of the grand potential).
From Eqs.~(\ref{krmtd}) and (\ref{qp2}) we see that for chaotic systems the
integrand in Eq.~(\ref{grandk}) behaves as $\tau^{-3}$, while from
Eqs.~(\ref{kint}) and (\ref{qp2}) it follows that for integrable motion the
integrand varies as $\tau^{-4}$. Therefore in both cases the integral
(\ref{grandk}) converges for long times, even at zero temperature. The
dominant contributions come in fact from short times, where the divergence of
the integral is stopped by the cutoff at $\tau = \tau_{min}$ introduced by the
form factor. When applied to other thermodynamic quantities, the same
analysis shows that the fluctuations of the response function for integrable
and chaotic dynamics, and those of the particle number for integrable systems,
are also dominated by the short-time contributions. Because the short-time
dynamics is specific to each system, then in general the second moments of
those functions are non--universal, and consequently the same is true for the
probability distribution. This conclusion remains valid as the temperature
increases, since a finite temperature truncates the long orbits.

We thus conclude that {\sl the fluctuations of the grand potential, the energy
and the response function for integrable and chaotic dynamics, and those of
the particle number for integrable systems, are dominated, at an
arbitrary temperature, by the shortest non-universal periodic orbits of the
system}.

As mentioned before, the sum for $\langle \widetilde{\cal N}^2 \rangle$ is
convergent only for integrable systems. The analysis shows that the shortest
orbits are not providing the dominant contribution when considering the
fluctuations of the particle number of chaotic systems. The same conclusion is
reached at low temperatures for other thermodynamic functions, like the
entropy or the specific heat. As $T \rightarrow 0$ the moments of the
distribution of the fluctuations of these quantities pick up contributions
from orbits with period $\tau_p \geqa \tauh$, whose statistical behavior is
universal. Conversely, in the energy domain the thermodynamic fluctuations of
these quantities therefore depend on the statistical properties of the
single-particle dynamics on a scale $\leq \delta$. To see this explicitly in a
particular case, consider the second moment of the entropy, given from
Eq.(\ref{grandk}) and Table~I by
\begin{equation} \label{vars}
\langle \widetilde{S}^2 \rangle = \frac{\hbar^2}{2 \pi^2}
\int_0^{\infty} \frac{d\tau}{\tau^4} ~ K(\tau) ~ \katp~\!\!\!^2 (\tau) \ .
\end{equation}
The function $\katp (\tau)$ has a peak centered at $\tau \approx \tauc$ of
height proportional to $1/T$ (for $\tau/\tauc \rightarrow 0$ it vanishes as
$\katp (\tau) \approx - (\tau/\tauc)^2 /3 T$). In the limit $T \rightarrow 0$
this function therefore selects in the integral (\ref{vars}) arbitrarily long
orbits. The variance of the entropy is therefore dominated by the long
universal orbits whenever $\tauc \gg \tau_{min}$ or, equivalently, for
temperatures $\kb T \ll E_c /2 \pi^2$. On the contrary, for temperatures $\kb
T \approx E_c /2 \pi^2$ non-universal orbits in the form factor are selected
and the result will depend on the system-specific peculiarities of the short
orbits. The explicit dependence of $\langle \widetilde{S}^2 \rangle$ on
temperature will be computed below. Similar arguments may be applied to other
quantities, like the specific heat. We therefore conclude that those functions
have, at $\kb T \ll E_c /2 \pi^2$, universal fluctuation distributions whose
functional form depend only on the universality class of the local
single--particle fluctuations on a scale $\leqa \delta$.

\section{The second moment in the $\tau_{min}$--approximation} \label{4}

There is in fact a simple way to estimate the variance in all the cases
considered above, independently of the short or long-time origin of the
dominant contributions. The approximation consists in using in
Eq.~(\ref{grandk}) the appropriate statistical form factor (given by
Eqs.~(\ref{kint}) and (\ref{krmt}) for integrable and chaotic motion,
respectively), the corresponding value of $Q^2 (\tau)$ from Eq.~(\ref{qp2}),
and to put $K(\tau)=0$ for $\tau < \tau_{min}$ in all cases. This is clearly
an approximation, that we call the $\tau_{min}$-approximation, since we are
extrapolating the statistical behavior of the orbits down to times $\tau
\approx \tau_{min}$, ignoring the short-time system-dependent structures. All
the short-time structures are condensed into a single parameter, the period of
the shortest orbit. For the fluctuations of the thermodynamic quantities
dominated by the shortest orbits -- whose second moment is described more
accurately below, see Eq.~(\ref{grandiag}) -- this is clearly a rough
approximation. On the contrary, for the entropy at low temperatures, and for
the variance of the particle number in chaotic systems, the leading order will
be well described by this approximation while the error, as we shall now see,
is made in the correction terms.

The virtue of the $\tau_{min}$-approximation is to provide simple estimates of
the size of the fluctuations, as well as of their dependence with chemical
potential, particle number, or external parameter. It requires a
minimum amount of information of the system, and is thus of interest in the
analysis of experiments.

\subsection{Zero temperature.} \label{4a}

Setting $\kat = 1$ and $\katp = 0$ in Eq.(\ref{grandk}), in this approximation
the results are expressed in terms of $\tau_{min}$, the parameters $\alpha$
and $\omega$ entering in Eq.(\ref{qp2}), the shell strength $g$ defined in
Section \ref{2c}, and the symmetry parameter $\beta$ in chaotic systems. Since
the integrals obtained from Eq.(\ref{grandk}) are straightforward, we do not
give here a detailed account of their computation. The results obtained are
summarized in Table~II. For chaotic systems with time reversal symmetry
$(\beta=1)$, only the leading order terms in $1/g$ are included. The first row
of the table also describes the variance of the grand potential.

\begin{center}
\begin{tabular}{||c||c|c|c||}
\hline\hline & & & \\ ~~~ & Integrable & Chaotic $\beta=1$ & Chaotic $\beta=2$
\\ \hline\hline & & & \\ $\langle \widetilde{U}^2 \rangle$ & $\displaystyle
\frac{1}{24 \pi^4} g E_c^2$ & $\displaystyle \frac{1}{8 \pi^4} E_c^2 \left(
1-\frac{2}{g}+{\cal O} (g^{-2} \log g) \right)$ & $\displaystyle \frac{1}{16
\pi^4} E_c^2 \left( 1 - \frac{1}{3 g^2} \right)$ \\ \hline & & & \\ $\langle
\widetilde{\cal N}^2 \rangle$ & $\displaystyle\frac{1}{2 \pi^2} \ g$ &
$\displaystyle \frac{1}{\pi^2} \left( \log g + 1 - \frac{\pi^2}{8} + {\cal O}
(g^{-1}) \right)$ & $\displaystyle \frac{1}{2 \pi^2} \left( \log g + 1
\right)$ \\ \hline & & & \\ ~~~ $\langle \widetilde{R}^2 \rangle$ ~~~ & ~~~
$\displaystyle \frac{1}{2 \pi^2} \ \omega \ g$ ~~~ & ~~~ $\displaystyle
\frac{1}{\pi^2}\frac{ \alpha}{\tau_{min}} \left( 1 - \frac{\log g}{g} + {\cal
O} (g^{-1}) \right) $ ~~~ & ~~~ $\displaystyle \frac{1}{2
\pi^2}\frac{\alpha}{\tau_{min}} \left( 1 - \frac{1}{2 g} \right)$ ~~~ \\
\hline\hline
\end{tabular}
\vspace{0.5cm} 

{\bf Table~II}: Zero-temperature second moments obtained in
the $\tau_{min}$-approximation
\end{center}

From these results it is possible to establish the scaling of the variances
with the Fermi energy and with the system size $L$ for a gas contained in a
$d$--dimensional cavity. We assume here that the parameter $x$ entering in the
definition of the response is a geometrical parameter that controls the shape
of the cavity (other parameters can be treated likewise). Since $\tau_{min}
\propto L \ef^{-1/2}$, then $E_c \propto L^{-1} \ef^{1/2}$. From
Eqs.~(\ref{nb}) and (\ref{mls}) (generalized to $d$--dimensions) we have
$\delta^{-1} \propto L^d \ef^{d/2-1}$. It then follows that $g \propto (L
\ef^{1/2})^{d-1}$. It has moreover been shown \cite{ls,lm1} that for shape
deformations of a cavity $\alpha \propto L^{-1} \ef^{3/2}$, and $\omega
\propto L^{-2} \ef^{2}$. Using these relations, we obtain the results
summarized in Table~III. The dependence with the number of fermions $N$ in the
gas has been computed assuming a constant density $n=N/V$ of the gas.
Therefore $\ef$ remains constant and, according to Eq.~(\ref{nb}), the size of
the system increases like $L \propto N^{1/d}$ with the number of particles.
Different dependencies are obtained when the volume is kept fixed. It is also
interesting to compare the typical quantum fluctuations to the magnitude of
the average part. For instance, since at constant density $\overline{U}
\propto N$ (for any dimension) (cf \cite{landau}), then $\sqrt{\langle
\widetilde{U}^2 \rangle} / \overline{U} \propto 1/N^{(d+3)/2 d}$.

\begin {center}
\begin{tabular}{||c||c|c||}
\hline\hline && \\ ~~~ & ~~~~~ Integrable ~~~~~ & ~~~~~ Chaotic ~~~~~ \\
\hline && \\ ~~~~~ $\langle \widetilde{U}^2 \rangle$ ~~~~~ & $L^{d-3}
\ef^{(d+1)/2} \sim N^{(d-3)/d}$ & $L^{-2} \ef \sim N^{-2/d}$ \\ \hline && \\
$\langle \widetilde{\cal N}^2 \rangle$ & $(L \ef^{1/2})^{d-1} \sim
N^{(d-1)/d}$ & $\log(L \ef^{1/2}) \sim \log N$ \\ \hline && \\ $\langle
\widetilde{R}^2 \rangle$ & $L^{d-3} \ef^{(d+3)/2} \sim N^{(d-3)/d}$ & $L^{-2}
\ef^{2} \sim N^{-2/d}$ \\ \hline\hline
\end{tabular}
\vspace{0.5cm}

{\bf Table~III}: Zero--temperature scaling of the variances with the system
size $L$, Fermi energy, and with the number of particles $N$ in the gas
in a $d$-dimensional cavity (keeping the density fixed).
\end{center}

\subsection{Temperature dependence of the fluctuations} \label{4b}

The variances depend on temperature through the function $\kat$ (or its
derivative), contained in the function $F(\tau)$ in Eq.(\ref{grandk}). It would
be too cumbersome to go through each function in detail. To be specific, we
illustrate the results by computing the temperature dependence of the second
moment of the entropy fluctuations, Eq.(\ref{vars}). In the
$\tau_{min}$--approximation, Eq.(\ref{vars}) is written
\begin{equation} \label{s0}
\langle \widetilde{S}^2 \rangle = \frac{\kb^2}{2} \int_{0}^\infty 
\frac{d x}{x^4} \left( \frac{x}{\sinh x} - \frac{x^2 \cosh x}{\sinh^2 x} 
\right)^2 K(x,\xh) ,
\end{equation}
where $K(x,\xh)=(1/\tauc) K(\tau=x \tauc)$, $\xh=\tauh/\tauc$, and
$K(x,\xh)=0$ for $x<x_{min}$, where $x_{min}=\tau_{min}/\tauc$. 

\begin{figure}
\begin{center}
\leavevmode
\epsfysize=2.6in
\epsfbox{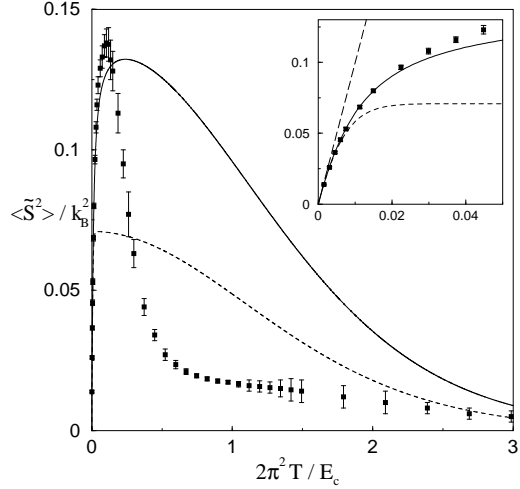}
\end{center}
%\vspace{0.5cm}
\caption{Temperature dependence of the entropy fluctuations in chaotic
systems, with $g=130$. Full and dashed lines: Eq.~(\ref{s0})
($\tau_{min}$--approximation) for $\beta=1$ and $\beta=2$, respectively.
Squares: Fermi gas in a Sinai cavity. The inset shows the behavior close to
the origin. The long-dashed straight line corresponds to Eq.(\ref{s4}).}
\label{s2c}
\end{figure}

In {\sl chaotic} systems three different temperature regimes can be
distinguished.\\

* {\sl Low temperatures: } $\kb T \ll \delta/2 \pi^2$. 
In this regime the function $\katp (\tau)$ is centered at times much larger
than $\tauh$. We therefore ignore all the structure of the form factor at
finite $\tau$ and take the asymptotic limit $K (x,\xh) \approx \xh$ in
Eq.(\ref{s0}). Then
\begin{equation} \label{s2}
\langle \widetilde{S}^2 \rangle = \frac{\kb^2 \xh}{2} \int_{0}^\infty \frac{d
x}{x^4} \left( \frac{x}{\sinh x} - \frac{x^2 \cosh x}{\sinh^2 x} \right)^2 \ .
\end{equation}
Denoting
\begin{equation} \label{s3}
I_k = \int_0^\infty \frac{d x}{x^k} 
\left( \frac{x}{\sinh x} - \frac{x^2 \cosh x}{\sinh^2 x} \right)^2 \ ,
\end{equation}
the variance of the entropy in the low temperature regime grows like
\begin{equation} \label{s4}
\langle \widetilde{S}^2 \rangle = \frac{1}{2} \ I_4 \ \kb^2 \ 
\frac{\kb T}{(\delta/ 2\pi^2)} \ ,
\end{equation}
where $I_4 \approx 0.153842$. The variance therefore increases linearly with
temperature, with a slope proportional to the average single-particle density
of states. It is interesting to remark that, up to a constant prefactor, the
low--temperature growth of the quantum fluctuations agrees with that of the
thermal fluctuations of the gas, which are of completely different physical
origin \cite{landau}. It is also interesting to compare the typical size of
the quantum fluctuations to their average part. From Eq.(\ref{s4}) we have
$\sqrt{\langle \widetilde{S}^2 \rangle} = \sqrt{I_4 /2} \ \kb \sqrt{\kb T /
(\delta/ 2\pi^2)}$, to be compared to the (low--temperature) growth of the
average part $\overline{S} = (1/6) \kb (\kb T / (\delta/ 2\pi^2))$. \\

* {\sl Intermediate temperatures: } $\delta/2\pi^2 \ll \kb T \ll E_c/2 \pi^2$.
In this regime the form factor may be approximated by its linear behavior,
Eq.(\ref{krmtd}). Using moreover the $\tau_{min}$ approximation for the
integral we get from Eq.~(\ref{s0})
\begin{equation} \label{s5}
\langle \widetilde{S}^2 \rangle = \frac{1}{\beta} \ I_3 \ \kb^2 \ 
- \frac{1}{18 \beta} \kb^2 \left( \frac{\kb T}{E_c/ 2\pi^2} \right)^2 \ ,
\end{equation}
where $I_3 = 0.141832$. There is therefore a saturation of the magnitude of the
entropy fluctuations in this intermediate regime at the ``quantum of entropy''
$\sqrt{\langle \widetilde{S}^2 \rangle} \approx \kb$, with a weak negative
quadratic dependence on temperature. Notice also the inverse dependence on the 
symmetry parameter $\beta$, which is absent in Eq.~(\ref{s4}).\\

* {\sl High temperatures: } $\kb T \gg E_c/2 \pi^2$. Since the form factor is
different from zero for times $\tau \geq \tau_{min}$, in this limit we can
approximate the function $\katp (\tau)$ by its exponential tail $\katp (\tau)
\approx - (2/T) (\tau/\tauc)^2 \exp (-\tau/\tauc)$. Then, in the
$\tau_{min}$--approximation and using again the linear short-time behavior of
$K(\tau)$ we find
\begin{equation} \label{s6}
\langle \widetilde{S}^2 \rangle = \frac{1}{\beta} \ \kb^2 \ 
\left[ 1 + 2 \left( \frac{\kb T}{E_c/ 2\pi^2} \right) \right] 
{\rm e}^{- 2 \kb T/(E_c/2 \pi^2)}   \ .
\end{equation}
At high temperatures compared to $E_c$ we thus observe the expected
exponential decay of the quantum fluctuations due to the thermal smoothing.

The behavior of the entropy fluctuations as a function of temperature
computed from Eq.~(\ref{s0}) for chaotic systems in this approximation is
displayed in Fig.~\ref{s2c} for the two symmetry classes. 

\begin{figure}
\begin{center}
\leavevmode
\epsfysize=2.8in
\epsfbox{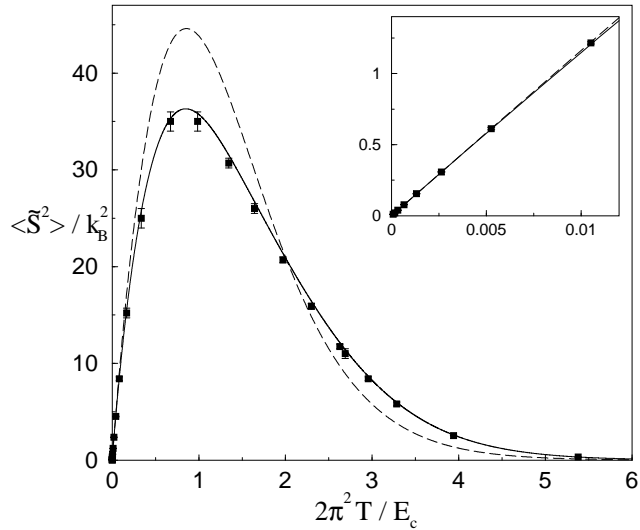}
\end{center}
%\vspace{0.5cm}
\caption{Temperature dependence of the entropy fluctuations in regular
systems, with $g=1500$. Long--dashed line: Eq.~(\ref{s0})
($\tau_{min}$--approximation). Squares: Fermi gas in a rectangular cavity.
Full line: semiclassical sum over periodic orbits, Eq.(\ref{grandiag}).
The inset shows the behavior close to the origin.}
\label{s2i}
\end{figure}

The situation is different in {\sl integrable systems} due to the different
behavior of the form factor, Eq.(\ref{kint}). Accordingly, there are only two
temperature regimes.\\

* {\sl Low temperatures: } $\kb T \ll E_c/2 \pi^2$. In this regime the form
factor is approximated by its asymptotic value $\tauh$, ignoring its
short-time structure. The result is identical to the low temperature regime of
chaotic systems, Eq.(\ref{s4}). The main difference with respect to the
chaotic case is that now this linear growth extends up to temperatures of
order $E_c /2\pi^2$ instead of $\delta /2\pi^2$. At this temperature, the typical entropy fluctuations
$\sqrt{\langle \widetilde{S}^2 \rangle}$ are of order $\sqrt{g \kb^2}$, which
are a factor $\sqrt{g} \gg 1$ greater than the maximum fluctuations in chaotic
systems.\\

* {\sl High temperatures: } $\kb T \gg E_c/2 \pi^2$. As in chaotic systems,
the function $\katp (\tau)$ is approximated by its exponential tail, and use
is made of the $\tau_{min}$ approximation for the form factor, $K(\tau) = 0$
for $\tau < \tau_{min}$ and $K(\tau) = \tauh$ for $\tau > \tau_{min}$. We get
\begin{equation} \label{s7}
\langle \widetilde{S}^2 \rangle = \kb^2 \ 
\left( \frac{\kb T}{\delta/ 2\pi^2} \right)  
{\rm e}^{- 2 \kb T/(E_c/2 \pi^2)}   \ .
\end{equation}
Fig.~\ref{s2i} shows, in this approximation, the behavior of the second moment
of the entropy fluctuations in integrable systems.

\begin{figure}
\begin{center}
\leavevmode
\epsfysize=1.3in
\epsfbox{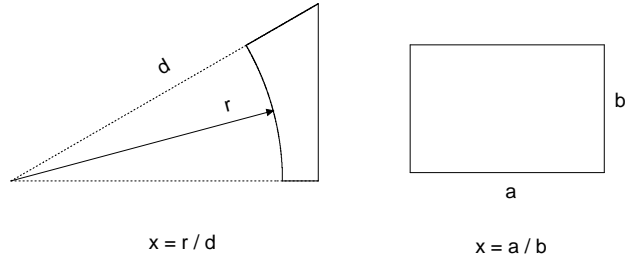}
\end{center}
%\vspace{0.5cm}
\caption{Cavities used to confined the Fermi gas. Left: Sinai (chaotic).
Right: rectangle (integrable). The parameter $x$ as defined in each case is
varied to modify the shape (while keeping the area constant).}
\label{bill}
\end{figure}

The second moment of the entropy fluctuations computed in the
$\tau_{min}$--approximation are expected to be accurate in the regime $2 \pi^2
\kb T \leqa \delta$, where only long orbits (compared to $\tau_{min}$)
contribute. This is confirmed by comparing with numerical simulations of a
Fermi gas in a cavity. Two different geometries are considered (cf
Fig.~\ref{bill}). In the first, a Fermi gas is contained inside a $2$D cavity
with infinite wall potential having a Sinai shape. The corresponding classical
single--particle motion is fully chaotic. The second is similar, but the
cavity has a rectangular shape. This guarantees an integrable single--particle
dynamics. The results obtained for the numerical simulations of a Fermi gas in
these two geometries are plotted in Figs.~\ref{s2c} and \ref{s2i} (squares). A
good agreement is observed indeed in the regime $2 \pi^2 \kb T \leqa \delta$
(cf the insets).

As temperature increases, shorter orbits become relevant for the entropy
fluctuations. For temperatures $2 \pi^2 \kb T \approx E_c$ or higher the
$\tau_{min}$--approximation fails because a statistical treatment of the
orbits is not appropriate any more. These deviations are clearly seen in
Figs.~\ref{s2c} and \ref{s2i}. In this range a precise description of the
decay of the fluctuations requires to explicitly take into account the
different short periodic orbits of the single-particle motion, which are
dominant. This treatment, which requires a much more detailed knowledge about
the system than the simple $\tau_{min}$--approximation, is considered in the
following section (\ref{5b}). It produces in the case of the rectangular
cavity the much more accurate description represented by the full curve in
Fig.~\ref{s2i}.

The temperature dependence of other thermodynamic quantities may be computed
analogously. For example, the second moment of the energy of chaotic systems
shows a quadratic decrease up to temperatures $\kb T \approx E_c/2 \pi^2$,
followed by an exponential tail for temperatures $\kb T \gg E_c/2 \pi^2$. A
qualitatively similar behavior is obtained for integrable systems. The
temperature dependence of the second moment of the particle number and of the
response function may be computed similarly.

\section{Higher moments and distributions} \label{5}

\subsection{Universal thermodynamic fluctuations} \label{5a}
 
In this case the fluctuations can be directly related to the local
universal statistics of the single--particle energy levels. The latter
correspond to an uncorrelated sequence for integrable systems, and to a random
matrix sequence in the chaotic case. This type of fluctuations are observed in
some thermodynamic functions at temperatures $\kb T \ll E_c/(2\pi^2)$.

The simplest quantity having a universal distribution at low temperatures is
the particle number in chaotic systems with a fixed chemical potential. It has
been conjectured that at zero temperature the fluctuations of the particle
number in chaotic systems are Gaussian distributed \cite{abs} (cf part (b) of
Fig~\ref{distc}). Since the variance of $\widetilde{\cal N}$ was already
computed in Section \ref{4}, the distribution is completely fixed in the range
$\kb T \ll E_c/2\pi^2$ where the universality holds.

Another thermodynamic function having universal fluctuations that we
consider here in detail is the entropy. When expressed in terms of the
single-particle energy levels $E_j$, this function, defined by Eq.(\ref{s}),
takes the form
\begin{equation} \label{s1}
S (x,\mu,T)= \kb \sum_j \log \left[ 1+e^{(\mu-E_j)/\kb T} \right] - 
\frac{1}{T} \sum_j \frac{\mu - E_j}{1+e^{(E_j - \mu)/\kb T}} \ . 
\end{equation}
At zero temperature, and as a function of the chemical potential, $S
(x,\mu,T)$ vanishes everywhere except when $\mu = E_j$, $j=1,2,\ldots$, where
it takes the value $S = \kb \log 2$. As temperature increases, peaks of width
proportional to $T$ centered at $\mu = E_j$ develop. At a given chemical
potential $\mu$, the entropy is therefore the superposition of the
contribution of the different peaks. In the range $\kb T \ll \delta/2\pi^2$
this superposition is only determined by the neighboring energy levels $E_j
\approx \mu$, and by their local statistical properties. We thus see, in a
direct manner, that at low temperatures the statistical properties of the
entropy depend only on the local statistics of the single-particle energy
levels. The second moment of that distribution as a function of temperature
was computed in Section \ref{4}.

\begin{figure}
\begin{center}
\leavevmode
\epsfysize=3.2in
\epsfbox{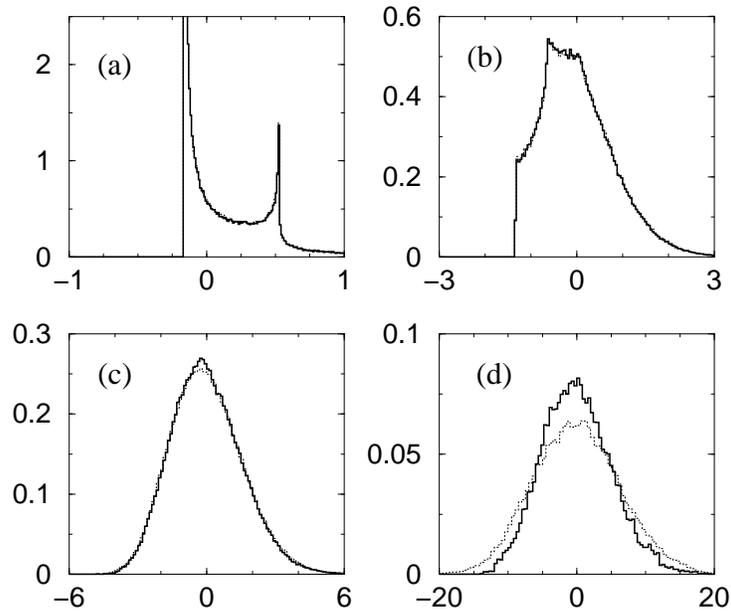}
\end{center}
%\vspace{0.2cm}
\caption{Histogram of the entropy probability distribution for integrable
single--particle motion at different temperatures. Dotted line: uncorrelated
spectrum. Full line: Fermi gas in a rectangular box with $x=(\sqrt{5}+1)/2$,
$\mu=(5.75\pm 0.75)\times 10^6$, and $g\approx 1500$. (a) $\kb T = \delta/(2
\pi^2)$; (b) $\kb T = 8\times \delta/(2 \pi^2)$; (c) $\kb T =
32\times\delta/(2 \pi^2)$; (d) $\kb T = 512\times \delta/(2 \pi^2)\approx
(1/3) E_c/(2 \pi^2)$. In parts (a), (b) and (c) the two curves are almost
indistinguishable.}
\label{sdisti}
\end{figure}

\begin{figure}
\begin{center}
\leavevmode
\epsfysize=3.2in
\epsfbox{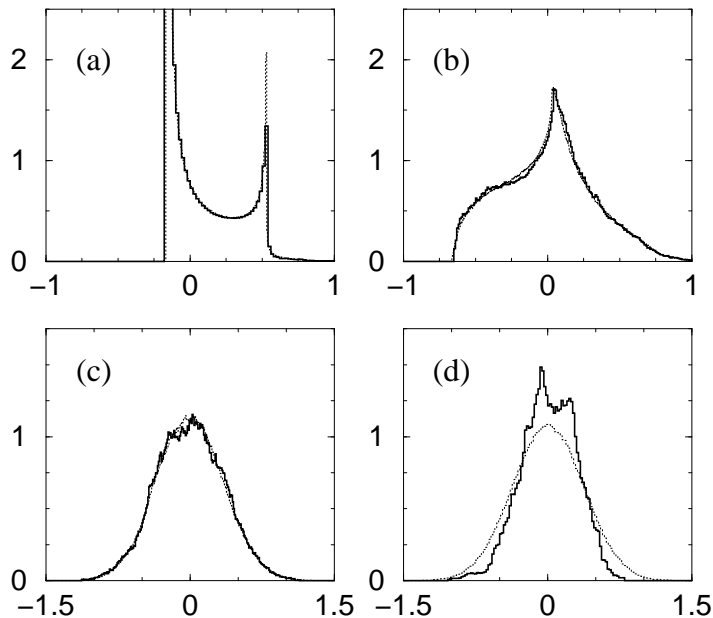}
\end{center}
%\vspace{0.2cm}
\caption{Histogram of the entropy probability distribution for chaotic
single--particle motion at different temperatures. Dotted line: random matrix
spectrum ($\beta=1$). Full line: Fermi gas in a Sinai box with $r=0.765 \pm
0.0025$, $\mu=900 \pm 100$, and $g\approx 130$. (a) $\kb T = \delta/(2
\pi^2)$; (b) $\kb T = 4\times \delta/(2 \pi^2)$; (c) $\kb T = 8\times\delta/(2
\pi^2)$; (d) $\kb T = 32 \times \delta/(2 \pi^2)\approx (1/4) E_c/(2 \pi^2)$.}
\label{sdistc}
\end{figure}

To compute the full distribution for regular and chaotic dynamics in the range
$\kb T \ll E_c/2 \pi^2$ one therefore has to compute the distribution of the
entropy $S$ from Eq.(\ref{s1}) by assuming an uncorrelated distribution and a
random matrix distribution for the $E_j$'s, respectively. We have done a
numerical computation of those probability distributions as a function of
temperature. The results obtained are summarized in Figs.~\ref{sdisti} and
\ref{sdistc}.

Figure \ref{sdisti} compares the distribution of the entropy fluctuations
obtained from Eq.~(\ref{s1}) when the single--particle energy levels $E_j$ are
assumed to be uncorrelated with the probability distributions computed for a
Fermi gas in a rectangular box. For temperatures up to $2 \pi^2 \kb T
\approx 30 \delta \approx E_c/50$ (part (a), (b), and (c)), the two
distributions are almost indistinguishable and show a high sensitivity to
temperature variations. For temperatures of the order $E_c /2\pi^2$ or higher
(part (d) of Fig.~\ref{sdisti}) the universality of the probability
distribution is lost and strong deviations are observed. The entropy
probability distribution for the uncorrelated spectrum tends to a Gaussian.

Figure \ref{sdistc} shows a similar comparison but for a chaotic
single--particle dynamics. It compares the probability distribution of the
entropy fluctuations computed from Eq.~(\ref{s1}) when the single--particle
energy levels $E_j$ are eigenvalues of a random matrix ensemble with
$\beta=1$, with the probability distribution computed for a Fermi gas in a
Sinai box. Again, a remarkable agreement is found for temperatures $2 \pi^2
\kb T \ll E_c$, and departures are observed as $E_c$ is approached. The entropy
probability distribution for the random matrix spectrum tends to a Gaussian.

For temperatures of order $E_c /2\pi^2$ or higher (part (d) in
Figs.~\ref{sdisti} and \ref{sdistc}), where the universality is lost, the
moments of the probability distributions, which are now determined by the
short periodic orbits, are non--universal and may be computed using the
techniques developed below. One should keep in mind that the scale $E_c$ and
the shell-strength parameter $g$, set in a real system by the inverse time of
flight across the system, are totally absent and therefore not defined in a
purely uncorrelated or random matrix spectrum.

\subsection{Non--universal thermodynamic fluctuations} \label{5b}

They occur when the fluctuations are dominated by the non-universal aspects of
the short time dynamics. This type of fluctuations are observed in all the
thermodynamic quantities for temperatures $\kb T \geq E_c /(2\pi^2)$. But they
also occur, as we saw in section \ref{3}, in certain thermodynamic functions
at arbitrary temperatures (including $T=0$). The $\tau_{min}$--approximation
used in Section \ref{4} to compute the second moment is not very accurate in
this case since that approximation is too crude. Here we provide accurate
expressions to compute all the moments of the distribution for non--universal
fluctuations.

Since off diagonal terms involving different short periodic orbits in products
like Eq.~(\ref{vargen2}) are killed by the average procedure, the dominant
contribution to the double sum will come from the rapidly convergent diagonal
terms $p = p'$. From Eq.~(\ref{vargen2}) the dominant diagonal contribution to
the second moment of thermodynamic quantities with non--universal
fluctuations, that we denote by a subscript ``$0$'', is
\begin{equation}\label{grandiag}
\langle \widetilde{\Phi}^2_0 \rangle = 2 \ {\cal C}^2 
\sum_{p,r} {\cal A}_{p,r}^2 \ . 
\end{equation}
The advantage of the $\tau_{min}$--approximation computed in the previous
section compared to the more accurate expression Eq.(\ref{grandiag}) is that
only information about the period of the shortest orbit is required, while
Eq.(\ref{grandiag}) involves much more detailed information about the
periodic orbits.

\begin{figure}
\begin{center}
\leavevmode
\epsfysize=2.6in
\epsfbox{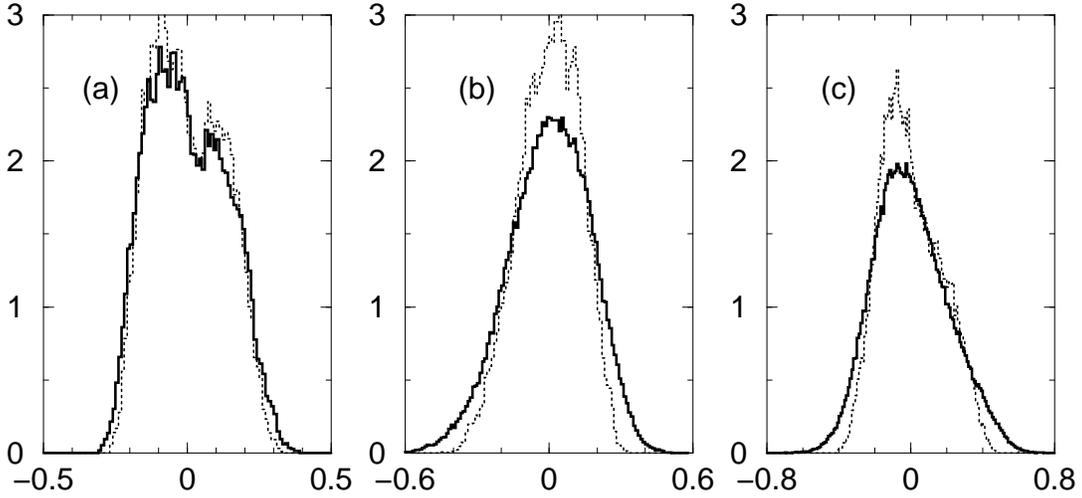}
\end{center}
%\vspace{0.5cm}
\caption{Histogram of the probability distribution of normalized thermodynamic
functions for a Fermi gas in a 2D rectangular box: (a)
$\widetilde{U}/\mu^{3/4}$, (b) $\widetilde{\cal N}/\mu^{1/4}$, (c)
$\widetilde{R}/\mu^{5/4}$. The distributions are computed at
$x=(1+\sqrt{5})/2$, in the window $\mu=(1 \pm 0.1)\times 10^8$ ($g=1500$).
Full line: $T=0$. Dashed line: $\kb T = 1273 \times \delta/(2\pi^2) \sim 0.5
\times E_c/(2 \pi^2)$.}
\label{disti}
\end{figure}

The diagonal approximation Eq.~(\ref{grandiag}) can be generalized to compute
all the moments of the oscillating quantities. Unlike the variance, a main
difference concerning the higher moments is that the repetitions of the
periodic orbits play now an essential role. We will explicitly go through the
calculation of the third and fourth moments, the generalization to the higher
ones being straightforward. We return to the notation of equation
(\ref{genosc}), keeping in mind the dependence on chemical potential,
parameter and temperature of the variables, but not writing it explicitly. By
the semiclassical formula the third power is written as a triple sum
\begin{equation}
\widetilde{\Phi}^3 = {\cal C}^3 \sum_{p_1,r_1} {\cal A}_{p_1,r_1} 2 \cos
\alpha_{p_1,r_1} \sum_{p_2,r_2} {\cal A}_{p_2,r_2} 2 \cos \alpha_{p_2,r_2}
\sum_{p_3,r_3} {\cal A}_{p_3,r_3} 2 \cos \alpha_{p_3,r_3} \ ,
\end{equation}
where $\alpha_{p_i,r_i}= r_i S_{p_i}/\hbar+\nu_{p_i,r_i}$. The product of
three cosines gives
\begin{eqnarray}
8 \cos \alpha_{p_1,r_1} \cos \alpha_{p_2,r_2} \cos \alpha_{p_3,r_3} &=& 2 \cos
(\alpha_{p_1,r_1}+\alpha_{p_2,r_2}+\alpha_{p_3,r_3}) + 2 \cos
(\alpha_{p_1,r_1}+\alpha_{p_2,r_2}-\alpha_{p_3,r_3}) \nonumber \\ &+& 2 \cos
(\alpha_{p_1,r_1}-\alpha_{p_2,r_2}+\alpha_{p_3,r_3}) + 2 \cos
(\alpha_{p_1,r_1}-\alpha_{p_2,r_2}-\alpha_{p_3,r_3}) \nonumber \ .
\end{eqnarray}
The average kills the first term in the r.h.s.. The last three terms give the
same result interchanging indices. Then
\begin{equation}
\langle \widetilde{\Phi}^3 \rangle = 6 ~ {\cal C}^3 \left \langle
\sum_{p_1,r_1} \sum_{p_2,r_2} \sum_{p_3,r_3} {\cal A}_{p_1,r_1} {\cal
A}_{p_2,r_2} {\cal A}_{p_3,r_3} \cos
(\alpha_{p_1,r_1}+\alpha_{p_2,r_2}-\alpha_{p_3,r_3}) \right \rangle _{\mu,x} \
. \label{m3a}
\end{equation}
The main dependence of this function on the chemical potential comes, as
usual, from the variations of the argument $\Delta \phi$ of the cosine. In the
averaging window it is enough to consider a linear approximation
\begin{equation}
\Delta \phi = \left( r_1 \tau_{p_1} + r_2 \tau_{p_2} - r_3 \tau_{p_3} \right)
~ \frac{\Delta \mu}{\hbar} \ .
\end{equation}
The wild oscillatory behavior produce by the phase variations will typically
have zero average, except for those terms for which $\Delta \phi = 0$.
Assuming incommensurability of the periods of all the orbits, the
only way to satisfy this condition is by imposing
\begin{equation}
\tau_{p_1} = \tau_{p_2} = \tau_{p_3} \ , ~ r_1 + r_2 = r_3 \label{diag3} \ . 
\end{equation}
This is a generalization of the diagonal approximation, and illustrates the
basic mechanism to calculate the higher moments. In the argument of the cosine
in Eq.~(\ref{m3a}) only the Maslov indices survive because the equality of the
periods implies the equality of the actions. The above constraints simplify
the multiple sums and finally a compact expression for the third moment is
obtained
\begin{equation}
\langle \widetilde{\Phi}^3_0 \rangle = 6 ~ {\cal C}^3 \sum_p
\sum_{r_1=1}^{\infty} \sum_{r_2=1}^{\infty} {\cal A}_{p,r_1} ~ {\cal
A}_{p,r_2} ~ {\cal A}_{p,r_1+r_2} \cos
(\nu_{p,r_1}+\nu_{p,r_2}-\nu_{p,r_1+r_2}) \ . \label{m3b}
\end{equation}

When the fluctuations are dominated by the short orbits, the convergence of
this series is guaranteed by the same arguments that showed the convergence of
the second moment. The sum (\ref{m3b}) is in general different from zero. This
produces a finite skew giving a non-symmetric probability distribution.

\footnotesize

\begin{center}
\begin{tabular}{||c||r|r||r|r||r|r||}
\hline\hline
~ & \multicolumn{2}{c||}{ ~ } & \multicolumn{2}{c||}{ ~ } &
\multicolumn{2}{c||}{ ~ } \\
$k$ & \multicolumn{2}{c||}{$\widetilde{U}/\mu^{3/4}$} &
\multicolumn{2}{c||}{ $ \widetilde{ \cal N} / \mu^{1/4}$ } &
\multicolumn{2}{c||}{ $ \widetilde{R} / \mu^{5/4}$} \\ 
\cline{2-7} \hline &&&&&& \\
~ & $T=0$ ~~~~ & $T_1 \neq 0$ ~~~~ & $T=0$ ~~~~ & $T_1 \neq 0$ ~~~~ & $T=0$ ~~~~ &
$T_1 \neq 0$ ~~~~ \\
\hline\hline &&&&&& \\
2 & $(1.81 \pm 0.01)\times 10^{-2}$ & $(1.55 \pm 0.01)\times 10^{-2}$ & $(2.83
\pm 0.05)\times 10^{-2}$ & $(1.63 \pm 0.03)\times 10^{-2}$ & $(4.25 \pm 0.10)
\times 10^{-2}$ & $(2.84 \pm 0.06) \times 10^{-2}$ \\
~ & $1.83290 \times 10^{-2}$ & $1.57853 \times 10^{-2}$ & $2.8477 \times
10^{-2}$ & $1.65251 \times 10^{-2}$ & $4.3148 \times 10^{-2}$ & $2.89043 \times
10^{-2}$ \\
\hline &&&&&& \\
3 & $(5.0 \pm 0.4) \times 10^{-4}$ & $(3.5 \pm 0.5) \times 10^{-4}$ & $-(1.3
\pm 0.2) \times 10^{-3}$ & $-(5.5 \pm 0.7) \times 10^{-4}$ & $(2.4 \pm 0.3)
\times 10^{-3}$ & $(1.6 \pm 0.1) \times 10^{-3}$ \\
~ & $5.31694 \times 10^{-4}$ & $3.81036 \times 10^{-4}$ & $-1.35453 \times
10^{-3}$ & $-5.85974 \times 10^{-4}$ & $2.44332 \times 10^{-3}$ & $1.58641
\times 10^{-3}$ \\
\hline &&&&&& \\
4 & $(7.1 \pm 0.2) \times 10^{-4}$ & $(4.95 \pm 0.05) \times 10^{-4}$ & $(2.3
\pm 0.1) \times 10^{-3}$ & $(6.8 \pm 0.3) \times 10^{-4}$ & $(5.0 \pm 0.3)
\times 10^{-3}$ & $(1.9 \pm 1.0) \times 10^{-3}$ \\
~ & $7.3133 \times 10^{-4}$ & $5.13214 \times 10^{-4}$ & $2.33807\times
10^{-3}$ & $7.14034 \times 10^{-4}$ & $5.16324 \times 10^{-3}$ & $1.99110
\times 10^{-3}$ \\
\hline\hline
\end{tabular}
\vspace{0.5cm}

{\bf Table~IV}: Moments of the distribution of some thermodynamic functions
with non--universal fluctuations for a Fermi gas in a rectangular cavity. In
each row, the upper values are the numerical results obtained from the
distributions shown in Fig.~\ref{disti}, while the lower values are obtained
from the semiclassical sums over periodic orbits. The temperature $T_1 =
\kb T = 1273 \times \delta/(2\pi^2)$ is the same as in Fig.~\ref{disti}.
\end{center}

\normalsize

The fourth moment is calculated in a similar way. The product of four cosines
gives eight terms. One of them, containing the sum of all the actions, has
zero average. The other terms, arranged after interchanging indices, give
\begin{eqnarray}
\langle \widetilde{\Phi}^4 \rangle &=& 2 ~ {\cal C}^4 \Big\langle
\sum_{p_1,r_1} \cdots \sum_{p_4,r_4} {\cal A}_{p_1,r_1} {\cal A}_{p_2,r_2}
{\cal A}_{p_3,r_3} {\cal A}_{p_4,r_4} \big[ ~ 4 ~ \cos
(\alpha_{p_1,r_1}+\alpha_{p_2,r_2}+\alpha_{p_3,r_3}-\alpha_{p_4,r_4})
\nonumber \\ & & ~~~ + ~ 3 ~ \cos
(\alpha_{p_1,r_1}+\alpha_{p_2,r_2}-\alpha_{p_3,r_3}-\alpha_{p_4,r_4}) ~ \big]
~ \Big\rangle \ .
\end{eqnarray}
We restrict again to those terms having exactly zero phase variation,
neglecting quasi cancellations of long orbits. Then for the first cosine in
the r.h.s. this is equivalent to $r_1 \tau_{p_1} + r_2 \tau_{p_2} + r_3
\tau_{p_3} - r_4 \tau_{p_4}=0$, and by the incommensurability of the periods
the only solution is
\begin{equation}
p_1 = p_2 = p_3 = p_4 \ , ~~ r_1 + r_2 + r_3 = r_4 \nonumber \ .
\end{equation}
As for the third moment, this condition gives a simple sum over primitive
periodic orbits. The three indices $r_1,r_2$ and $r_3$ are free, and $r_4$ is
determined by them.

For the second cosine the corresponding equation $r_1 \tau_{p_1} + r_2
\tau_{p_2} - r_3 \tau_{p_3} - r_4 \tau_{p_4}=0$ admits two different
solutions. The first one is
\begin{equation}
p_1 = p_2 = p_3 = p_4 \ , ~~ r_1 + r_2 = r_3 + r_4  \nonumber \ .
\end{equation}
The difference is that only $r_1$ and $r_2$ are free. Because the repetitions
are positive integer numbers, $r_3$ varies between 1 and $r_1+r_2-1$,
and $r_4$ is determined by the constraint.
The other solution, which has a weight two because of the possibility of
interchanging indices, is
\begin{equation}
p_3 = p_1 ~ , ~ p_4 = p_2 ~ , ~ r_3 = r_1 ~ , ~ r_4 = r_2
\nonumber \ .
\end{equation}
This reduces the expressions to double sums over two different primitive
periodic orbits, $p_1$ and $p_2$, and the repetitions $r_1,r_2$. To eliminate
the inequality between $p_1$ and $p_2$, we can add and subtract a term where
the two indices are equal. With this term the double sum, both in primitive
periodic orbits and repetitions, factorizes giving a term directly
proportional to the square of the second moment, in its diagonal
approximation. Finally we arrive at a convergent formula for the fourth moment
\begin{eqnarray}
\langle \widetilde{\Phi}^4_0 \rangle &=& 2 ~ {\cal C}^4 \sum_p \Big[ ~ 4
\sum_{r_1,r_2,r_3=1}^{\infty} {\cal A}_{p,r_1} ~ {\cal A}_{p,r_2} ~ {\cal
A}_{p,r_3} ~ {\cal A}_{p,r_1+r_2+r_3} \cos
(\nu_{p,r_1}+\nu_{p,r_2}+\nu_{p,r_3}-\nu_{p,r_1+r_2+r_3}) \nonumber \\ &~& ~~
+ ~ 3 \sum_{r_1,r_2=1}^{\infty} \sum_{r_3=1}^{r_1+r_2-1} {\cal A}_{p,r_1} ~
{\cal A}_{p,r_2} ~ {\cal A}_{p,r_3} ~ {\cal A}_{p,r_1+r_2-r_3} \cos
(\nu_{p,r_1}+\nu_{p,r_2}-\nu_{p,r_3}-\nu_{p,r_1+r_2-r_3}) \nonumber \\ &~& ~~
- ~ 6 \sum_{r_1,r_2=1}^{\infty} {\cal A}^2_{p,r_1} ~ {\cal A}^2_{p,r_2} \Big]
+ 3 ~ \langle \widetilde{\Phi}^2 \rangle^2 \ . \label{m4}
\end{eqnarray}
The last term $3\langle \widetilde{\Phi}^2 \rangle^2$ is the result for the
fourth moment if the distribution would have been Gaussian. The remaining terms
in Eq.~(\ref{m4}) produce deviations from that behavior (the excess of the
distribution). 

The generalization to higher moments is straightforward. The formulae are more
complex, but they are always written as a sum over one primitive periodic
orbit index, the repetitions originating from different constraints, plus terms
coming from the previous moments (e.g. for the fifth moment there is a term
proportional to$\langle \widetilde{\Phi}^2 \rangle \langle \widetilde{\Phi}^3
\rangle$).

\begin{figure}
\begin{center}
\leavevmode
\epsfysize=2.6in
\epsfbox{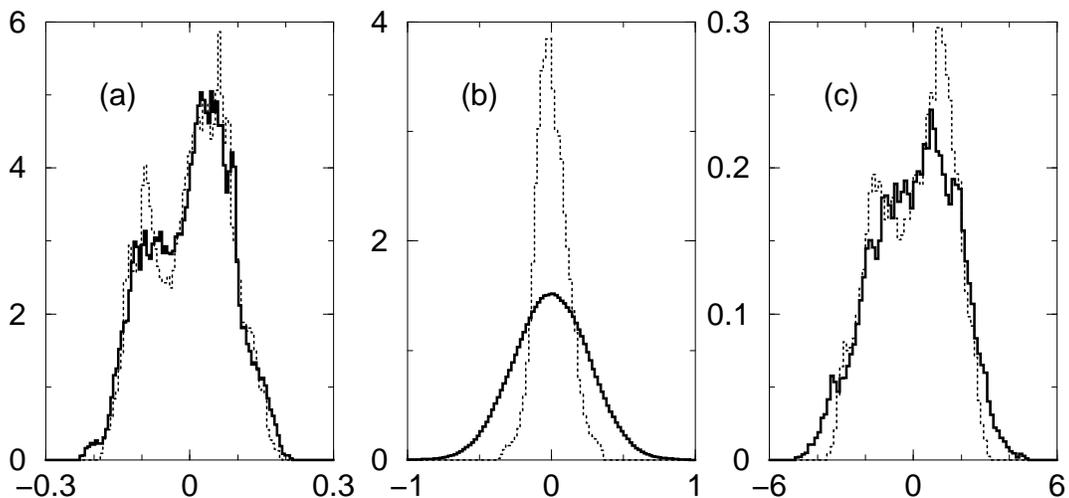}
\end{center}
%\vspace{0.5cm}
\caption{Histogram of the probability distribution of normalized thermodynamic
functions for a Fermi gas in a 2D Sinai box: (a) $\widetilde{U}/\mu^{3/4}$,
(b) $\widetilde{\cal N}/\mu^{1/4}$, (c) $\widetilde{R}/\mu^{5/4}$. The
distributions are computed in the windows $r=0.765 \pm 0.0006$, $\mu=750 \pm
250$. Full line: $T=0$. Dashed line: $\kb T = 0.2 \times E_c/(2 \pi^2)$.}
\label{distc}
\end{figure}

Figures \ref{disti} and \ref{distc} illustrate these results. They show the
distribution of the quantum fluctuations of the total energy, particle number
and response of a Fermi gas contained in a 2D rectangular and Sinai cavity,
respectively. The fluctuating part of the thermodynamic functions is
normalized by an appropriate power of the chemical potential. This
normalization eliminates their zero--temperature energy dependence (cf
Table~III with $d=2$). The probability distributions are normalized by setting
their area to one. The parameter $x$ used to compute the response controls the
shape of the cavity (cf Fig.~\ref{bill}), while the area is kept constant. $R$
is therefore a ``quantum'' force, related to the variations of the energy of
the gas produced by the quantum shell corrections.

The first four moments of the probability distributions shown in
Fig.~\ref{disti} have been computed. The numerical results obtained, at the
two different temperatures, are compared in Table~IV to the first four moments
calculated using the analytic expressions obtained above based on the periodic
orbits. The overall precision obtained is of the order of $2\%$ (in the worst
case), and is within the numerical uncertainties. For comparison, the rescaled
second moments of the energy, particle number and response computed at $T=0$
using the $\tau_{min}$--approximation are $1.74\times 10^{-2}$, $3.2\times
10^{-2}$ and $3.98 \times 10^{-2}$, respectively. They have an accuracy of
$4\%$, $13\%$ and $6\%$ with respect to the numerical values of Table~IV,
respectively.

We have also computed from Eq.~(\ref{grandiag}) the temperature dependence of
the variance of the entropy fluctuations for a Fermi gas in the rectangular
box using the shortest orbits. The result is represented by the full line in
Fig.~\ref{s2i}, which shows a very good agreement with respect to the
numerical values computed directly from the Fermi gas. Such a good agreement
is found even at temperatures of order $\delta$ because, in integrable
systems, the ``diagonal'' approximation used above is almost exact. This is
not expected to occur in chaotic systems. We were not able to make similar
comparisons for the chaotic Sinai cavity because we haven't computed the
periodic orbits for that system.

The fluctuations of the particle number in the chaotic case,
Fig.~\ref{distc}(b), shows at $T=0$ the universal Gaussian behavior discussed
in section V.A. At higher temperatures the universality is lost.
 
\section{Correlation functions} \label{6}

When the parameter $x$ or chemical potential $\mu$ are varied, the
autocorrelation function of a thermodynamic quantity $\Phi$ is defined as
\begin{equation}
C_{\Phi}(x,\mu) = \left \langle \widetilde{\Phi}
(x_0-x/2,\mu_0-\mu/2,T) ~ \widetilde{\Phi} (x_0+x/2,\mu_0+ \mu/2,T) \right 
\rangle _{x_0,\mu_0} \ . \label{corr1}
\end{equation}
In the following, we assume that the regular or chaotic nature of the
single--particle motion in the gas is unchanged when the parameters
$(x_0,\mu_0)$ vary. When $\widetilde{\Phi}$ is expressed as a sum over the
periodic orbits using Eq.(\ref{genosc}), $C_{\Phi}$ takes the form of a double
sum over periodic orbits labelled by the indices $p$ and $p'$. As usual, only
the term involving one half the cosine of the difference of the actions gives
a non-zero average. The amplitudes ${\cal A}_p$ and actions $S_p$ are
evaluated at $(\mu_0-\mu/2,x_0-x/2)$, while those with index $p'$ at
$(\mu_0+\mu/2,x_0+ x/2)$. In the semiclassical regime these functions will
have a slow variation in a quantum scale, and therefore to leading order the
parameter dependence can be ignore in the prefactors, and only a linear
expansion of the actions is taken
$$
S_p(\mu_0 \pm \frac{\mu}{2},x_0 \pm \frac{x}{2}) \approx S_p(\mu_0,x_0)
\pm \frac{\tau_p \mu}{2} \pm \frac{Q_p x}{2} \ .
$$  
The autocorrelation is now written
\begin{equation}
C_{\Phi}(x,\mu) = 2 ~ {\cal C}^2 ~ \left \langle \sum_{p,p'} \sum_{r,r'=1}^{\infty} {\cal A}_{p,r} {\cal A}_{p',r'} \cos \left[ \frac{r S_p
- r' S_{p'}}{\hbar} + \frac{(r \tau_p + r' \tau_{p'})}{2 \hbar} \mu +
\frac{(r Q_p + r' Q_{p'})}{2 \hbar} x \right] \right \rangle \ ,
\label{cint}
\end{equation}
where all the classical functions are evaluated at $(x_0,\mu_0)$. The
average over a small window $\Delta x$ around $x_0$ restricts the sums to
orbits having approximately the same $Q_p$, while the energy average imposes
the same condition over the periods.

To proceed further there are two main paths, corresponding to universal and
non--universal fluctuations. For the latter, since short orbits dominate, the
diagonal approximation furnishes the main contribution to the autocorrelation,
\begin{equation}
C_{\Phi,0}(x,\mu) = 2 ~ {\cal C}^2 ~ \sum_p \sum_{r=1}^{\infty}
{\cal A}_{p,r}^2 \cos \left( \frac{r ~ Q_p ~ x}{\hbar} + \frac{r ~
\tau_p ~ \mu}{\hbar} \right) \ .
\end{equation}
In this equation the shortest periodic orbits have the largest prefactors
${\cal A}_{p,r}$ and the lowest frequencies inside the cosine, giving rise in
general to an erratic and non-decaying oscillation as a function of the
parameters. The typical period of these oscillations is dominated by the
shortest periodic orbit. The correlation lengths, both in chemical potential
and external parameter $x$ are estimated making the phase variation equal to
$2 \pi$ giving
\begin{equation}
\mu^* = E_c = \frac{h}{\tau_{min}}  ~~~~~~~~~ x^* = \frac{h}{|Q_{min}|} 
\ .
\end{equation}

In the presence of universal fluctuations, the treatment of Eq.(\ref{cint}) is
different. Since long orbits give an important contribution, the diagonal
approximation is not sufficient and a more accurate evaluation is needed.
General expressions for the autocorrelation of the gas were given in
Ref.\cite{lm1}. Ordering the orbits by their period, and taking into account
the restrictions imposed by the averaging procedure, the autocorrelation of a
thermodynamic quantity $\Phi$ takes the form
\begin{equation} \label{corr2}
C_{\Phi}(x,\mu) = \frac{{\cal C}^2}{2 \pi^2 \hbar^2}
\int_0^{\infty} \frac{d\tau}{\tau^4}~ \cos (\mu \tau/\hbar) ~ K(\tau,x) ~ 
F (\tau) \ .
\end{equation}
The constant ${\cal C}$ and the function $F (\tau)$ are defined as in
Eq.(\ref{grandk}), to which Eq.(\ref{corr2}) reduces when $x=\mu=0$
($K(\tau,x=0) = K(\tau))$. The function $K(\tau,x)$ is a generalization of the
form factor: it is the Fourier transform of the parametric two-point
correlation function $\langle \tilde{\rho} (\mu_0 - \mu/2, x_0 - x/2)
\tilde{\rho} (\mu_0 + \mu/2, x_0 + x/2) \rangle_{\mu_0}$ with respect to
$\mu$. Semiclassically its expression is given by Eq.(\ref{form}) with the
replacement $\cos \left[(S_p-S_{p'})/\hbar \right] \rightarrow \cos
\left[(S_p-S_{p'})/\hbar + (Q_p+Q_{p'})/2\hbar \right]$. For chaotic systems
and for times $\tau \gg \tau_{min}$, $K(\tau,x)$ has been conjectured to be a
universal function depending only on the symmetry parameter $\beta$ and
described by random matrix theory \cite{sa1}.

In many cases this integral may lead to results which are determined only by
the universal, long time aspects of the dynamics. Consider for example the
autocorrelation of the entropy. Then $F(\tau)=\katp \ \!\!\!^2 (\tau)$ in
Eq.(\ref{corr2}). At temperatures $\kb T \ll E_c/2 \pi^2$ the function $\katp
\ \!\!\!^2 (\tau)$ selects only times much longer than $\tau_{min}$. Then the
integral in (\ref{corr2}) is insensitive to the system-dependent features of
$K(\tau,x)$ present at short times, and will therefore take a universal form
depending only on the temperature, the mean level spacing and the typical
slope of the single-particle energy levels at Fermi energy \cite{sa1,ls}. For
(fully) chaotic systems we therefore need the random matrix theory result
for this function. Changing for simplicity to the scaled variable
$t=\tau/\tauh$, using the definition $K(t,x)=(1/\tauh) K(\tau=\tauh~t,x)$ and
scaling the parameter $x$ according to the prescription given in \cite{ls},
for chaotic systems without time reversal symmetry for instance ($\beta = 2$)
we can Fourier transform with respect to energy the parametric two-point
correlation function obtained in \cite{sa1}. We get
\begin{equation}
K_{rmt}(t,x) = \left\{ \matrix{ e^{-2 \pi^2 x^2 t} \sinh(2 \pi^2 x^2 t^2)/2
\pi^2 x^2 t \; \; \; \; \; & t \leq 1 \cr \!\!\!\!\!\!\!\!\! 
e^{-2 \pi^2 x^2 t^2} \sinh(2 \pi^2
x^2 t)/2 \pi^2 x^2 t & \ t > 1 \ . \cr } \right.
\end{equation}
Inserting this in Eq.(\ref{corr2}) the autocorrelation of the entropy may
therefore be computed for temperatures well below $E_c/2\pi^2$. 

\section{Discussion} \label{7}

It follows from Eqs.(\ref{nb}) and (\ref{mls}) that the three "intrinsic"
energy scales that characterize the Fermi gas, namely $\ef$, $E_c$ and
$\delta$, are determined by the number of fermions in the gas (taking into
account degeneracies due to, e.g., spin), by their mass, and by the volume of
the cavity (or, alternatively, $\mu$, $m$, and $V$ fix ${\overline N}$, $E_c$
and $\delta$). The same is true for the adimensional shell--strength parameter
$g=E_c/\delta$. They are independent of the precise shape of the system, and
are therefore independent of the regular or chaotic nature of the
single--particle dynamics.

\begin {center}
\begin{tabular}{||c||c|c|||}
\hline\hline && \\ ~~~ & ~~~~~ Metal particles ~~~~~ & ~~~~~ Nuclei ~~~~~ 
%\\ && 
\\ \hline && \\ $N$ & $20$ -- $5000$ & $25$ -- $250$ %\\ && 
\\ \hline && \\ $r=r_0 N^{1/3}$ & $0.5$ -- $3.5$ nm & $3.2$ -- $7$ fm %\\ && 
\\ \hline && \\ ~~~~~ $n = N/V$ ~~~~~ & $30$ electrons nm$^{-3}$ & 
$0.17$ nucleons fm$^{-3}$ %\\ && 
\\ \hline && \\ $k_F r$ & 10 -- 67 & 11 -- 25 %\\ && 
\\ \hline && \\ $\ef$ & $35000$ K & $37$ MeV %\\ && 
\\ \hline && \\ $E_c = \displaystyle \frac{\pi \ef}{\kf r}$ & $22000$ -- $3500$
K & $26.5$ -- $12.3$ MeV %\\ && 
\\ \hline && \\ $\delta = \displaystyle \frac{2 \ef}{3 N}$ & $1200$ -- $5$ K & 
$1$ -- $0.1$ MeV %\\ && 
\\ \hline && \\ $g = \displaystyle \frac{3 \pi N}{2 \kf r}$ & $20$ -- $760$ & 
$27$ -- $125$ %\\ && 
\\ \hline\hline
\end{tabular}
\vspace{0.5cm}

{\bf Table~V}: Main features of metal particles and atomic nuclei
as a function of particle number, \\ ranging from $20$ to $5000$ and from $25$
to $250$, respectively.
\end{center}

It is useful to recall their typical magnitude as well as other relevant
physical parameters for two important systems, metallic particles and atomic
nuclei. Table~V summarizes the main features. The number of particles is $N$,
electrons or nucleons respectively. It typically ranges from $20$ to several
thousands in metal particles, and from $25$ to $250$ in nuclei. For
definiteness, we take the window 20--5000 for metallic particles (although
values up to $20000$ can be reached experimentally \cite{martin}). The values
indicated in the table are for these two windows, respectively. The radius
$r=r_0 N^{1/3}$ fixes the system size as a function of $N$. For metal
particles $r_0$ (which is sometimes denoted $r_s$) ranges from $1$ to $3
\AA$. In the table we have used the typical value $r_s = 2 \AA$. For nuclei
$r_0 = 1.1$ fm.

Other relevant experimental systems, with much larger typical sizes, are the
two--dimensional (2D) quantum dots. For instance, the electron gases contained
in microscopic 2D cavities created in high mobility GaAs heterojunctions have
a typical electronic density of $n \approx 5\times 10^{11}$ cm$^{-2}$ and
Fermi energy $E_F = 18$ meV (see, e.g., Ref.\cite{bh}). For typical
quantum--dot sizes $L \approx 0.1$ -- 5 $\mu$m, the other characteristic
energy scales vary in the range $E_c \approx 6.4$ -- $0.128$ meV and $\delta
\approx 0.36$ -- $1.44\times 10^{-4}$ meV. This gives an effective number of
fermions in the last shell in the range $g = 18$ -- $890$.

The table illustrates the variations in the energy scales with the number of
particles. $E_c$ is comparable to $\ef$ for small grains and nuclei. As the
system grows, the three energy scales become however well separated, thus
improving the applicability of the semiclassical methods.

As we have shown in the previous sections, the parameter $g$ which, according
to Table~V, may take quite large values in real systems, plays an important
role in setting the scale of the fluctuations of thermodynamic functions. As
summarized in Table~II, it determines the relative value of the fluctuations
in integrable systems as compared to chaotic ones for quantities like the
energy or the response of the gas. This amplification factor is due to the
special organization of periodic orbits in regular systems, which form
one--parameter families, to be compared to the isolated character of periodic
orbits for chaotic motion. Among the chaotic systems the variance is twice
bigger in systems with time reversal symmetry.

It is instructive to compare the fluctuations of the energy and of the
response function of the whole Fermi gas with the corresponding fluctuations
of a single--particle energy level located in the neighborhood of the Fermi
energy. When an external parameter is varied, the fluctuations of the energy
of a single--particle level located at $E \approx \ef$ may be computed from the
fluctuations of the Fermi energy of a system with a fixed number of particles
$N$. The number of particles is related to the Fermi energy through the
spectral counting function, $N = {\cal N} (\ef)$. Writing $\ef =
\overline{E}_F + \widetilde{E}_F$ and also splitting ${\cal N}$ into its
smooth and oscillatory parts, expanding when appropriate by assuming
$\widetilde{E}_F \ll \overline{E}_F$, and ignoring second order terms in the
fluctuations we arrive at $N = \overline{\cal N} (\overline{E}_F) +
\overline{\rho} (\overline{E}_F) \widetilde{E}_F + \widetilde{\cal N}
(\overline{E}_F)$. But $\overline{E}_F$ is precisely defined in order to
satisfy $\overline{\cal N} (\overline{E}_F) = N$. The fluctuations of a
single--particle level, estimated by $\widetilde{E}_{sp} \approx
\widetilde{E}_F$, are therefore
\begin{equation}\label{sp1}
\widetilde{E}_{sp} = - \widetilde{\cal N} (\overline{E}_F) \ \delta
\end{equation}
with the variance 
\begin{equation}\label{sp2}
\langle \widetilde{E}^2_{sp} \rangle = \langle \widetilde{\cal N}^2 \rangle \ 
\delta^2 \ .
\end{equation}
This relation allows, together with the variance of the particle number
summarized in Table~II, to connect the energy fluctuations of the gas with
those of the single particle levels. To leading order in $g$ we get
\begin{equation}\label{sp3} 
\langle \widetilde{U}^2 \rangle = \left\{ \begin{array}{ll} \displaystyle
\frac{1}{12 \pi^2}
g^2 \langle \widetilde{E}^2_{sp} \rangle \;\;\;\;\;\; & \mbox{integrable} \; ,
\\ & \\ \displaystyle \frac{1}{8 \pi^2} \frac{g^2}{\log g} \langle
\widetilde{E}^2_{sp} \rangle & \mbox{chaotic} \; .
\end{array}\right.
\end{equation}

To establish a similar connexion for the response function we must first
determine the local fluctuations of the response of the single--particle
energy levels
\begin{equation}\label{rsp1} 
R_{sp} = \frac{\partial E_j}{\partial x} \ ,
\end{equation}
where $E_j \approx \ef$. It was shown in Ref.\cite{ls} that in chaotic systems
the variance of $R_{sp}$ (with respect to a mean which is subtracted) is
\begin{equation}\label{rsp2}
\langle \widetilde{R}^2_{sp} \rangle = \frac{2 \alpha}{\beta \tauh} =
\frac{2 \alpha}{\beta \tau_{min}}\frac{1}{g} \ ,
\end{equation}
where $\alpha$ was defined in Eq.(\ref{qp2}). A similar computation for
integrable systems gives, using the corresponding relation in Eq.(\ref{qp2})
\begin{equation}\label{rsp3}
\langle \widetilde{R}^2_{sp} \rangle = \omega.
\end{equation}
Using these relations together with the results for the variance of the
response of the gas $\langle \widetilde{R}^2 \rangle$ from Table~II we find
\begin{equation}\label{rsp4}
\langle \widetilde{R}^2 \rangle = \frac{1}{2 \pi^2} g 
\langle \widetilde{R}^2_{sp} \rangle
\end{equation}
for both regular and chaotic dynamics.

The global picture emerging from these relations is quite instructive.
Contrary to naive expectations, Eq.(\ref{sp2}) shows that the typical
fluctuations of a single--particle level located at $\ef$ are not of order
$\delta$, but are "amplified" by the fluctuations of the particle number.
Since the latter are larger in integrable systems compared to chaotic ones (by
a relative factor $\sqrt{g/\log g}$ ), then $\sqrt{\langle
\widetilde{E}^2_{sp} \rangle_{int}/\langle \widetilde{E}^2_{sp}\rangle_{ch}}
\propto \sqrt{g/\log g}$. Eq.(\ref{sp3}) shows, moreover, that the
fluctuations of the energy of the gas are, in turn, much larger than the local
variations. Similar considerations hold for the response. We see from
Eq.(\ref{rsp4}) that the typical response of the gas, $R_{typ} = \sqrt{\langle
\widetilde{R}^2 \rangle}$, is $\sqrt{g}$ times greater than that of the
single--particle levels located at $\ef$, irrespective of the integrable or
chaotic nature of the dynamics. One should however keep in mind that $\langle
\widetilde{R}^2_{sp} \rangle$ is bigger in integrable systems with respect to
chaotic ones by a factor of order $g$, and therefore the overall fluctuations
of the response of the gas are more important if the dynamics is integrable.
When computed in the particular case of the persistent currents in mesoscopic
rings, $I = - c \ \partial \Omega/\partial \phi$, where the external parameter
$x$ is now the magnetic flux threading the ring, the formula $\langle
\widetilde{R}^2 \rangle = \alpha /(\pi^2 \tau_{min})$ in Table~II valid for
chaotic systems gives for the typical current $I_{typ} = \sqrt{\langle
\widetilde{I}^2 \rangle} = (\sqrt{2} c/\pi) E_c/\phi_0$ (where $c$ is the
speed of light and $\phi_0$ the flux quantum), in agreement with the result
obtained in \cite{crg} using a Green's--function approach, and with the
$\sqrt{g}$ amplification obtained for integrable systems (see also \cite{ag}).

All this results clearly show the close connexions existing between the
fluctuations of the gas and the parameter $g$. Remember that $g=E_c/\delta$
is the number of fermions contained in the last shell, i.e. in a window of
size $E_c$ below the Fermi energy. This number of particles has to be compared
with the total number of fermions $N=(2/d) \ef/\delta$ (in $d$ dimensions).
Eqs.~(\ref{sp3}) and (\ref{rsp4}) show that neither the total number of
particles, nor an individual single--particle level (located at
$\ef$), are responsible for the fluctuations of the gas, but rather the
intermediate number $g$, with typically $1 \ll g \ll N$. The quantum
fluctuations of the gas therefore originate from the contributions of the $g$
particles located in the last shell, as already pointed out in
Ref.~\cite{bdjpsw}. However, the way these particles contribute to the
fluctuations depend on the quantity considered. Eq.~(\ref{sp3}) shows that
they contribute ``coherently'' in the case of the total ground--state energy,
since $U_{typ} = \sqrt{\langle \widetilde{U}^2\rangle}$ is proportional to $g$
(ignoring the $\log g$ denominator in the chaotic case). In contrast, their
contributions to the response add up ``incoherently'' (as in a random walk),
as shown by Eq.~(\ref{rsp4}). Physically, this may be interpreted as the
independence of the slopes $\partial E_j/\partial x$ of the last $g$
single--particle levels.

The situation is more subtle for quantities like the entropy. As was shown in
section \ref{4b}, the temperature dependence of the variance for $\kb T \ll
\delta/2 \pi^2$ is unique, independent of the nature of the single--particle
dynamics (and given by Eq.~(\ref{s4})). However, at higher temperatures
differences are observed. The maximum of the typical value of the entropy
fluctuations, $S_{typ} = \sqrt{\langle \widetilde{S}^2 \rangle}$, is of order
$\kb$ for chaotic systems, and it is reached at $2 \pi^2 \kb T \approx
\delta$, whereas for integrable systems $S_{typ}$ grows up to values $\propto
g \kb$. The maximum value in integrable systems is reached at temperatures $2
\pi^2 \kb T \approx E_c$.

There are several distinct applications of the present results. Among them, we
mentioned the study of the electronic contribution to the mechanical force in
experiments where metallic nanocontacts are pulled \cite{nano}, and the
fluctuations of the energy of nuclei as a function of the number of nucleons.
Progress made in both directions will be published elsewhere. In a related
work, the distribution of the energy of a Fermi gas whose single--particle
energy levels are given by the imaginary part of the complex zeros of the
Riemann zeta function was recently investigated \cite{lmb}. This fictitious
fermionic system was named "the Riemannium". Aside its mathematical interest,
it serves as an explicit and important test of our results relating the
probability distribution of thermodynamic functions to periodic orbit theory.
This is because there exist strong connexions between the quantum theory of
chaotic systems and the complex zeros of the Riemann zeta function interpreted
as spectral eigenvalues. A high precision agreement was found between the
numerical results and the moments of the probability distribution of
$\widetilde{\Omega}$ computed from the results of Section \ref{5a} for the
Riemannium.

\section{Concluding remarks} \label{8}

Our results illustrate the strong connexions and deep relationships that exist
between, on the one hand, the thermodynamic quantum fluctuations of confined
Fermi gases and, on the other, the statistical theories of level fluctuations
and periodic orbit theory.

The regular or chaotic nature of the single--particle motion imprints the
single--particle spectrum in two different ways. The first one occurs at the
scale of the mean level spacing $\delta$, and produces different (but
universal) statistical fluctuations. The second acts on a much larger scale
$E_c$, set by the inverse of the time of flight across the system. A bunching
of the single--particle levels produced by the short periodic orbits is
observed on this scale, whose intensity depends on the regular or chaotic
nature of the motion.

The main theme of our investigations has been to determine how these two
features of the single--particle spectrum influence the probability
distribution of the quantum fluctuations of the different thermodynamic
functions of the gas, and to establish their temperature dependence. We find a
rich variety of phenomena and of different regimes. In some cases the
thermodynamic quantum fluctuations directly reflect the universality of the
single--particle spectrum on the scale $\delta$. Their corresponding
probability distributions are universal functions, whose shape depends on the
quantity and temperature considered, and on the regular or chaotic nature of
the dynamics, but not on the precise shape or other specific properties of the
system. In other cases, the fluctuations are, in contrast, totally determined
by the long--range modulations of the single--particle spectrum on scales
$E_c$. They are therefore insensitive to the universality present on the scale
$\delta$. For some thermodynamic quantities, this type of fluctuations are
dominant even at zero temperature. The shape of the distributions is, in this
case, system--dependent and therefore non--universal. The moments may be
computed from the short periodic orbits of the single--particle dynamics. At
temperatures of order $E_c$ or higher, all the thermodynamic fluctuations,
even those dominated at low temperatures by the universal features of the
single--particle spectrum, fall in this second class. In all cases and for the
different regimes, we have shown how the probability distributions can be
explicitly computed in each case.

The variance of the thermodynamic functions has also been investigated using a
simple approximation, that takes only into account the regular or chaotic
nature of the single--particle dynamics and the time of flight across the
system. It has the advantage of providing a good qualitative description for
the different thermodynamic quantities as a function of temperature, number of
particles, etc, while requiring a minimum amount of information about the
system. Although more accurate descriptions can be made if more information is
available, this approximation is clearly of interest in experiments where the
parameters controlling the system, like its shape, are not well known.

A closely related problem is the motion of quasiparticles in potentials with
bulk disorder. This case has been extensively treated in condensed matter
physics in particular as a model for the electronic properties of metals and
semiconductors. Compared to the ballistic motion considered here, the disorder
introduces a new scale, the elastic mean free path $\ell_e$ between
impurities. When $\ell_e$ is much smaller than the system size $L$ (and $L$
is much smaller than the localization length), the single--particle motion
across the system is diffusive. The corresponding energy scale is the Thouless
energy, $E_c=h D/L^2$, with $D$ the diffusion coefficient. Similarly to the
influence of the (short) periodic orbits in the ballistic case, the diffusive
motion produces long--range correlations in the single--particle spectrum
\cite{as}. However, the nature of the correlations are different in both
cases. Therefore, the statistics of the quantum thermodynamic fluctuations
that are dominated by the long range correlations are going to be different.
An analogous study for diffusive systems to the one made here in the ballistic
case is therefore necessary. In contrast, it has been shown that the local
fluctuations of the single--particle levels on a scale $\delta$ are, in the
metallic regime of disordered systems, also described by random matrix theory
\cite{efetov2}. Therefore, our results concerning the quantum fluctuations of
thermodynamic quantities in the universal regime apply also to disordered
systems.

The quantum fluctuations considered may be compared to other fluctuations, the
thermal ones. Although the latter are inherent to any thermodynamical
treatment, their physical origin and their experimental manifestation is very
different from that of the quantum fluctuations here considered.

Many experiments have verified the validity and accuracy of the physical
picture obtained from mean--field, single--particle approximations to
many--body systems. However the nature of the modifications induced by the
(residual) interactions in the probability distribution of the thermodynamic
fluctuations remains an open problem.\\

{\bf Acknowledgements:} we are in debt with C. Schmit who kindly bring us the
numerical spectrum of the Sinai cavity.

%%%%%%%%%%%%%%%%%%%%%%%%%%%%%%%

\end{document}